\documentclass[preprint]{elsarticle}
\usepackage[margin=1in]{geometry}
\usepackage[utf8]{inputenc}
\usepackage{cmap}
\usepackage{textcomp}
\usepackage{microtype}
\usepackage{subfiles}
\usepackage[numbers]{natbib}
\usepackage{mathtools}
\usepackage{graphicx}
\usepackage{xurl}
\usepackage{varioref}
\usepackage{hyperref}
\usepackage{cleveref}
\usepackage{xcolor}
\usepackage{color,soul}
\usepackage{siunitx}
\usepackage{float}
\usepackage{tikz}
\usepackage{pgfgantt}
\usepackage{pgfplots}
\usepackage{multirow}
\usepackage{multicol}
\usepackage{booktabs}
\usepackage{subcaption}
\usepackage{amsmath}
\usepackage{amsopn}
\usepackage{amssymb}
\usepackage[inkscapeformat=png]{svg}
\usetikzlibrary{positioning, arrows.meta, decorations.markings}

\input{glyphtounicode} % these 2 lines make it possible to parse information easily DO NOT REMOVE
\pgfplotsset{compat=1.16}

\newcommand{\markweighting}[1]
{%
}

\newcommand{\markexplanation}[1]
{%
}

\newcommand{\norm}[1]{\left\lVert#1\right\rVert}

\DeclareMathOperator{\E}{\mathbb{E}}
\newcommand{\resultsscale}{0.9}

\begin{document}
\begin{frontmatter}
    \title{%
        {Generative Deep Learning and Signal Processing for Data Augmentation of Cardiac Auscultation Signals: Improving Model Robustness Using Synthetic Audio}
    }

    \author[inst1]{Leigh Abbott}
    \affiliation[inst1]{organization={School of Electrical Engineering, Computing, and Mathematical Sciences (EECMS), Faculty of Science and Engineering, Curtin University},
    Department and Organization
        city={Bentley},
        postcode={6102}, 
        state={WA},
        country={Australia}
    }
    \author[inst1]{Milan Marocchi}
    \author[inst1]{Matthew Fynn}
    \author[inst1]{Yue Rong}
    \author[inst1]{Sven Nordholm}

    \begin{abstract}
        Accurately interpreting cardiac auscultation signals is essential for diagnosing and managing cardiovascular diseases. However, the paucity of labelled data inhibits classification models' training. Researchers have turned to generative deep learning techniques alongside signal processing to augment existing data and improve cardiac auscultation classification models.
        However, the primary focus of prior studies has been on model performance rather than robustness. Robustness, in this case, is defined as both in-distribution and out-of-distribution performance by measures such as Matthew’s correlation coefficient. 
        {One contribution of this work is to show} that more robust abnormal heart sound classifiers can be trained using an
            augmented dataset.
        The augmented dataset includes both signal processing techniques and synthetic phonocardiograms conditionally generated using the WaveGrad and DiffWave diffusion models an approach that, to the best of our knowledge, is the first work of its kind.
        {The efficacy of the proposed data augmentation approach is evaluated on an example convolutional neural network, trained on the original and augmented data.}
        It is found that both the in-distribution and out-of-distribution performance can be
            improved over various datasets {for neural networks trained} with this
            augmented dataset.
        Results show significant performance improvements. Specifically, in-distribution accuracy, balanced accuracy, and Matthew’s correlation coefficient (MCC) increased by 2.5\%, 4.1\%, and 0.066, respectively. The greatest out-of-distribution improvements were observed on one dataset, where accuracy, balanced accuracy, and MCC increased by 43.1\%, 20.2\%, and 0.297, respectively.
        These improvements across all metrics highlight that augmented datasets significantly address issues of imbalanced data, ultimately leading to more generalisable and robust classifiers.
    \end{abstract}


\end{frontmatter}

\section{Introduction}

Cardiovascular disease (CVD) is the primary contributor to mortality worldwide, representing more
    than \SI{30}{\percent} of all global deaths in 2019~\cite{who_cvd}.
In addition to the human cost, CVD places an immense economic burden on healthcare systems and
    society~\cite{who_cvd}.
To treat CVD effectively, it is necessary to diagnose and evaluate the condition of the heart accurately.

Cardiac auscultation (CA) is the process of listening to sounds generated by the
    heart~\cite{reed_heart_2004}.
Physicians have traditionally performed CA using stethoscopes to detect and monitor heart
    conditions in a non-invasive manner.
However, the difficulty of performing CA leads to uncertainty in diagnosis and poor patient
    outcomes.
The issue is further complicated by the fact that CA is both difficult to teach and a specialised skill, with studies noting that primary care physicians often lack proficiency in this area~\cite{reed_heart_2004}.

Recently, a wearable multichannel electrophonocardiography (EPCG) device has been
    developed~\cite{preprint_curtin}.
The premise of this device is to detect CVD utilising synchroised phonocardiogram (PCG) and electrocardiogram (ECG) data. The combination of these signals can result in more accurate and robust classifications.
However, there is currently limited synchronised multichannel phonocardiogram and electrocardiogram (SMPECG) data, which creates a need for a technique to aid in creating a larger dataset.

There are current limitations that prevent robust classification results across multiple datasets.
These include a lack of quality data and unbalanced datasets, with most data having lots of background
    noise, resulting in a low signal-to-noise ratio.
There is also a limited amount of synchronised PCG and ECG recordings, 
    which limits the effectiveness of algorithms, despite the large amounts of standalone ECG and some PCG data.
Traditional augmentation approaches can help to overcome these issues, with augmentation being
    applied to existing signals~\cite{cinc_augment_2016}.
This is somewhat lacking, however, as it does not always increase the out-of-distribution
    performance,  leaving room for further approaches to address this issue.
With recent advancements in conditional waveform generation using diffusion
    models~\cite{diffwave,wavegrad}, it is possible to extend previously ECG-only datasets by
    generating PCG signals conditioned from the ECG in these datasets.

This work explores traditional augmentation approaches alongside the generation of synthetic signals,
    to create more robust classifiers of abnormal heart sounds.

The main contributions of this work are summarised below:
\begin{itemize}
    \item Development of a diffusion model to create PCG signals conditional on existing ECG signals, allowing additional data to be used from ECG datasets once the diffusion model has created the corresponding PCG signal. To the best of our knowledge, this is the first work using diffusion models to generate PCG signals.
    \item Traditional augmentation methods synchronised over the PCG and ECG signals and extensive methods beyond those utilised in other studies.
    % SHOULD INCLUDE MORE METRICS HERE AT LEAST THE MCC.
    \item Augmentation methods were applied to a top-performing model~\cite{milan} on the training-a dataset~\cite{cinc_nov16}, resulting in improvements of 2.5\% in accuracy, 4.1\% in balanced accuracy, 1.9\% in $F_1^+$ score, and 0.066 in Matthew's Correlation Coefficient (MCC). 
    Additionally, when tested on the training-e dataset—where the model had not been trained on any of the dataset's data—there were notable improvements of 43.1\% in accuracy, 20.2\% in balanced accuracy, 27.1\% in $F_1^+$ score, and 0.297 in MCC. 
\end{itemize}

The remainder of the paper is organised as follows. Background in PCG and ECG {signals is covered in Section 2. Literature survey on model robustness, biomedical signal augmentation and generative models is presented in Section 3.}
Following this, the methods and results are presented in Sections 4 and 5 before a discussion of the results in Section 6 and the
    final conclusions and further work are summarised in Section 7.

\section{Background}

\subsection{Phonocardiogram and Electrocardiogram Signals}

PCG signals comprise multiple sounds from the opening and closing of valves and blood flow inside the heart that cause
    vibrations, which are then recorded from the chest wall~\cite{leatham}.
The fundamental heart sounds are the first (S1) and second (S2) sounds, which are the most prominent.
The S1 occurs during the beginning of the systole and is caused by isovolumetric ventricular
    contraction.
S2 is caused by the closing of the aortic and pulmonic valves during the beginning of the diastole.
Although the S1 and S2 sounds are the most audible, PCG signals consist of many other heart sounds
    such as the third (S3) and fourth (S4) heart sounds, systolic ejection clicks, mid-systolic
    clicks, opening snap and heart murmurs~\cite{cinc_nov16}.
These heart murmurs are produced by turbulent flowing blood, which can indicate the presence of particular CVDs.
These various heart sounds all lie within the low frequencies, with S1 from
    \SIrange{10}{140}{\hertz} and the highest energy around \SIrange{25}{45}{\hertz}.
The S2 is from \SIrange{10}{200}{\hertz}, with most of the energy around \SIrange{55}{75}{\hertz}.
S3 and S4 sounds are from \SIrange{20}{70}{\hertz}, although they are much less audible, mainly occurring in children and pathological subjects.
Murmurs are usually found in slightly higher frequencies and range from \SI{25}{\hertz} to
    \SI{400}{\hertz}~\cite{schmidt}, with some being found in frequencies higher than
    \SI{600}{\hertz}, but with far less energy~\cite{springer}.

ECG signals represent the heart's electrical activity~\cite{heart4}.
An ECG signal consists of the P, QRS complex, and T waves, with a U wave also
    occasionally present~\cite{heart5}.
These waves can contain information to aid in CVD diagnosis.
ECG signals are commonly filtered between \SI{0.5}{\hertz} and \SI{40}{\hertz} to remove baseline
    wander and unwanted noise and interference~\cite{ecgfiltering}.
For example, in the case of coronary artery disease patients, studies have documented that symptoms
    such as T-wave inversion, ST-T abnormalities, left ventricular hypertrophy, and premature
    ventricular contractions can be observed~\cite{ecgcad}.

Combining these two signals has produced superior results compared to classification using a single
    signal~\cite{milan}, suggesting that relevant features for classification exist within both signals.
The increase in performance suggests that utilising synchronised PCG and ECG data will help to
    create more accurate and robust classifiers.

\section{{Literature Survey}}
\subsection{Model Robustness}

\citeauthor{plex_2022} (2022) \cite{plex_2022} presented a state-of-the-art framework for
    enhancing model reliability, focusing on robust generalisation.
Robust generalisation allows a model to perform well on data outside the training
    set~\cite{plex_2022}, encompassing in-distribution (ID) and out-of-distribution (OOD)
    generalisation~\cite{plex_2022}.

ID generalisation pertains to a model's performance on data within the training distribution but
    outside the training set, addressing underfitting and overfitting
    issues~\cite{plex_2022, bias_variance_tradeoff}.
OOD generalisation, on the other hand, concerns a model's ability to handle data distributions
    different from the training set, addressing distribution shifts such as subpopulation shifts,
    covariate shifts, and domain shifts~\cite{plex_2022,ood_robust_2022}.

Perturbation resilience is the ability of a model to handle atypical and significantly different
    data, including corruption, distortion, artifacts, missing data, gaps, spectral masking, extreme
    noise, and defective inputs, which is critical in clinical settings.

\subsubsection{Measuring Model Robustness}

\Cref{tab:tradMeas} shows formulas for traditional binary
    classification performance measures derived from the confusion matrix in
    \Cref{fig:confusion_matrix}\cite{systematic_class_measures_2009,perf_scores_class_2022,survey_heart_sound_2023}.
Sensitivity (recall/true positive rate) and specificity (true negative rate) measure correct
    classifications of positive and negative cases,
    respectively~\cite{systematic_class_measures_2009}.
Precision (positive predictive value) and negative predictive value measures correctly classified
    positive and negative cases among classified cases,
    respectively~\cite{systematic_class_measures_2009}.
Accuracy measures overall correct classifications~\cite{systematic_class_measures_2009}.
Ideally, all these measures are unity, indicating no false predictions.

\begin{figure}
    [htbp] \centering 
\renewcommand{\arraystretch}{1.5}
\begin{tabular}{l|l|c|c|} % chktex 44
    \multicolumn{2}{c}{} & \multicolumn{2}{c}{Actual} \\
    \cline{3-4}
    \multicolumn{2}{c|}{} & Positive & Negative \\
    \cline{2-4}
    \multirow{2}{*}{Classified} & Positive & TP & FP \\
    \cline{2-4}
                                & Negative & FN & TN \\
                                \cline{2-4}
\end{tabular}
 \caption{Confusion
        Matrix}\label{fig:confusion_matrix}
\end{figure}

While having one target metric is ideal, it is impractical as each metric contains different
    information and no single measure captures all the information from a confusion
    matrix~\cite{perf_scores_class_2022}.
Summary metrics can be biased under certain conditions; for instance, accuracy can be misleading for
    imbalanced datasets.
Matthew's correlation coefficient (MCC) is a better single metric for classifier performance than F
    scores~\cite{mcc}.

\begin{table}[htbp]
        \caption{Traditional Measures}\label{tab:tradMeas}
        \centering
        \renewcommand{\arraystretch}{1.3}
        \begin{tabular}{lr@{~=~}l}
    \toprule
    \textbf{Metric} & \multicolumn{2}{l}{\textbf{Formula}} \\
    \midrule
    Sensitivity & $\text{TPR}$ & $\frac{{\text{TP}}}{{\text{TP} + \text{FN}}}$ \\
    Specificity & $\text{TNR}$ & $\frac{{\text{TN}}}{{\text{TN} + \text{FP}}}$ \\
    Precision & $\text{PPV}$ & $\frac{\text{TP}}{\text{TP} + \text{FP}}$ \\
    Negative Predictive Value & $\text{NPV}$ & $\frac{\text{TN}}{\text{TN} + \text{FN}}$ \\
    Accuracy & $\text{acc}$ & $\frac{{\text{TP} + \text{TN}}}{{\text{TP} + \text{TN} + \text{FP} + \text{FN}}}$ \\
    Balanced Accuracy & $\text{acc}_\mu$ & $\frac{\text{TPR}+\text{TNR}}{2}$ \\
    F1-Positive-Score & $\text{F}^+_1$ & $\frac{{2 \cdot \text{PPV} \cdot \text{TPR}}}{{\text{PPV} + \text{TPR}}}$ \\
    F1-Negative-Score & $\text{F}^-_1$ & $\frac{{2 \cdot \text{NPV} \cdot \text{TNR}}}{{\text{PNV} + \text{TNR}}}$ \\
    Matthew's Correlation Coefficient & MCC & $\frac{\text{TP} \cdot \text{TN} - \text{FP} \cdot \text{FN}}{\sqrt{(\text{TP}+\text{FP})(\text{TP}+\text{FN})(\text{TN}+\text{FP})(\text{TN}+\text{FN})}}$ \\
    \bottomrule
\end{tabular}

\end{table}

This work focuses on ID and OOD performance as the metric for model robustness, 
   focusing on balanced accuracy and MCC in addition to accuracy to present an overall indicator of the performance of the
    classification model.

\subsubsection{Model Robustness and Augmentation}

Data augmentation creates new data from existing data to increase the training set's size and
    variety, typically improving model performance.
To improve ID generalisation, providing more training data from the same distribution as the
    original data helps the model generalise to similar examples~\cite{plex_2022}.
To enhance OOD generalisation, extending the training data distribution beyond the original dataset,
    such as by balancing labels or adding scarce feature combinations, helps the model handle
    distribution shifts more effectively~\cite{ood_distshift_2022}.

\subsection{Generative Models}
\label{sec:tri}
Generative models are trained to learn the underlying distribution of the data to generate new
    samples.
As such, the goal is to train a mapping between the latent space and the data space so that the
    resulting samples are similar to the original data.
One of the important properties of the latent space is that it can enable the creation of new data
    through the manipulation of semantic representations of features and labels.
In recent history, three classes of models have advanced the field of generative learning in waves.

These classes are Autoencoders (AEs), Generative Adversarial Networks (GANs) and Diffusion models
    (DMs).
The first class of models, AEs, encode input data to a lower-dimensional latent space and then
    decode it back to the data space, often used in denoising models due to their ability to
    reconstruct the input from the latent space~\cite{autoencoders_2021}.
Variational Autoencoders (VAEs), an extension of AEs, regularise the latent distribution, enabling
    meaningful sampling from the latent space and removing discontinuities, thus facilitating
    generative capabilities~\cite{vaes_2019}.
GANs, the second class, consist of a generator and a discriminator network; the generator creates
    realistic samples from random noise, while the discriminator attempts to distinguish between
    real and synthetic samples, engaging in a zero-sum game to improve both
    networks~\cite{gans_seminal_2014}.
DMs, the third class, add random noise to input data and then train the model to reverse this
    process, learning to denoise data in a structured manner, with models like Latent Diffusion
    Models (LDMs) performing diffusion in the latent space for computational
    efficiency~\cite{diffusion_seminal_2015, diffusion_survey, ldms_seminal_2022}.

\begin{figure}[htbp]
    \centering
    \begin{tikzpicture}[auto,node distance=2.5cm]

    \usetikzlibrary{calc}
    \usetikzlibrary{positioning}

    \newcommand{\convexpath}[2]{%
    [
        create hullnodes/.code={%
            \global\edef\namelist{#1}
            \foreach [count=\counter] \nodename in \namelist {% chktex 1
                \global\edef\numberofnodes{\counter}
                \node at (\nodename) [draw=none,name=hullnode\counter] {}; % chktex 1
            }
            \node at (hullnode\numberofnodes) [name=hullnode0,draw=none] {}; % chktex 1
            \pgfmathtruncatemacro\lastnumber{\numberofnodes+1}
            \node at (hullnode1) [name=hullnode\lastnumber,draw=none] {}; % chktex 1
        },
        create hullnodes
    ]
    ($(hullnode1)!#2!-90:(hullnode0)$)
    \foreach [ % chktex 1
        evaluate=\currentnode as \previousnode using \currentnode-1, % chktex 1
        evaluate=\currentnode as \nextnode using \currentnode+1 % chktex 1
        ] \currentnode in {1,\numberofnodes} {% chktex 1 chktex 11
    -- ($(hullnode\currentnode)!#2!-90:(hullnode\previousnode)$)
      let \p1 = ($(hullnode\currentnode)!#2!-90:(hullnode\previousnode) - (hullnode\currentnode)$),
        \n1 = {atan2(\y1,\x1)},
        \p2 = ($(hullnode\currentnode)!#2!90:(hullnode\nextnode) - (hullnode\currentnode)$),
        \n2 = {atan2(\y2,\x2)},
        \n{delta} = {-Mod(\n1-\n2,360)} % chktex 1 chktex 11 chktex 36
      in
        {arc [start angle=\n1, delta angle=\n{delta}, radius=#2]}
    }
    -- cycle
    }

    \tikzset{%
        block/.style= {circle, draw, minimum size=2cm, align=center, text width=1.8cm, inner sep=0pt},
        arrow/.style= {thick, ->, >=stealth},
        line/.style= {thick, -},
        dotline/.style={thick, dash dot, draw=#1},
        dotline/.default=black,
        label/.style= {align=center, text width=2.8cm, inner sep=0pt},
    }

    \node [block] (HQ) {Sample Speed};
    \node [block] at (-120:3) (FS) {Sample Quality};
    \node [block] at (-60:3) (MCD) {Sample Variety};
    
    \node at (-120:1.5) (tempR) {};
    \node [label, above left of=tempR, red, anchor=north] (tempRR) {Generative Adversarial Networks};
    \node at (-60:1.5) (tempB) {};
    \node [label, above right of=tempB, blue, anchor=north] (tempBB) {Variational Autoencoders};
    \node at (-90:1.5) (tempG) {};
    \node [label, below of=tempG, green, anchor=north] (tempGG) {Diffusion Models};

    \draw[dotline,red] \convexpath{HQ, FS}{1.18cm};
    \draw[dotline,blue] \convexpath{MCD, HQ}{1.16cm};
    \draw[dotline,green] \convexpath{FS, MCD}{1.14cm};

\end{tikzpicture}
    \caption{The Generative Learning Trilemma}\label{fig:GLTrilemma}
\end{figure}

The ``generative learning trilemma'' may guide the trade-offs in choosing a generative learning
    model.
As \Cref{fig:GLTrilemma} (adapted from~\cite{gl_trilemma}) shows, models often excel at only two of
    three desired goals: high sample quality, fast sample speed, and large sample variety.
However, as mentioned earlier, performing the diffusion process in latent space allows LDMs to
    generate samples much faster, such that some argue it bypasses the trilemma in
    practice~\cite{ldms_seminal_2022, gl_trilemma}.
For this reason, LDMs have seen recent use in expanding datasets in biomedical projects, where data
    collection is prohibitively costly~\cite{brain_ldms}.
As such, this work aims to use both the WaveGrad and DiffWave diffusion models for the creation of PCG
    from ECG signals.

\subsection{Biomedical Signal Augmentation}

In~\cite{cinc_augment_2016}, data augmentation was employed to expand a PCG dataset from \SI{3153}{}
    recordings to \SI{53601} recordings, an increase by a factor of \SI{17}{}. \ The augmentation
    included a random combination of effects such as changes to pitch, speed, tempo, dither, volume,
    and mixing with audio~\cite{cinc_augment_2016}.
Despite achieving a sensitivity of \SI{96}{\percent} and a specificity of \SI{83}{\percent}, the
    authors concluded that their approach did not generalise well, with performance varying from
    \SI{99}{\percent} on the dataset with the most recordings to \SI{50}{\percent} on the dataset
    with the fewest recordings~\cite{cinc_augment_2016}.
Consequently, \citeauthor{cinc_augment_2016} \cite{cinc_augment_2016} suggested that more training data and further
    augmentation is necessary to enhance performance on unseen data. 

In a subsequent study by \citeauthor{analysis_data_aug_2022} \cite{analysis_data_aug_2022}, models trained with various
    augmentations were compared against a baseline.
Augmentations were applied to both the original and image-transformed data and were categorised by a
    ``physiological constraint'' 
    (whether the transform alters or violates physiological possibilities)
    and/or a 
    ``spectrogram constraint'' 
    (whether the transform alters the meaning of the spectrogram output)~\cite{analysis_data_aug_2022}.
Augmentations that violated the ``spectrogram constraint'' were linked to decreased model
    performance, while adherence to physiological possibilities was associated with improved
    performance~\cite{analysis_data_aug_2022}.
Notably, no single augmentation improved performance across all metrics, though some offered a more
    favorable trade-off than others~\cite{analysis_data_aug_2022}.

VAEs have been explored for the generation of synthetic lung auscultation sounds~\cite{lung_vaes_2022},
    where it was found that the use of VAE-generated signals in the training of classifiers were often
    improved, but not always, over training on just the original data.

GANs have also found lots of use within biomedical
    applications~\cite{lung_gans_2021,snore_gans_2020,heart_gans_2020}.
The introduction of synthetic data helps to overcome data imbalances as well as improve model
    performance.
In particular, GANs have been used to generate synthetic heart signals~\cite{heart_gans_2020}.
This work found that during early training, the waveform generated resembled a real signal
    with added noise~\cite{heart_gans_2020}.
Using the Empirical Wavelet Transform (EWT) to reduce this noise, the resulting signal at
    \SI{2000}{} epochs was more realistic than the resulting signal at \SI{12000}{} epochs, allowing
    for a sixfold reduction in training time~\cite{heart_gans_2020}.
Further work was performed to show that the generative model had not simply learned the training
    dataset~\cite{heart_gans_2020}.
As a result, the classifiers were able to classify the synthetic heart sounds correctly with
    accuracy greater than \SI{90}{\percent}~\cite{heart_gans_2020}.

In~\cite{general_biosig_gans_2023}, the general problem of generating synthetic one-dimensional
    biosignals are explored.
Both an autoencoder and GAN-based approach were explored.
To evaluate their models, the synthetic and real datasets are each used as either the training or
    test set for a classifier model that had previously achieved an accuracy of
    \SI{99}{\percent}~\cite{general_biosig_gans_2023}.
The results from this work showed that the synthetic data captured the underlying features and
    distributions of the real data and the synthetic data could be used to train classifiers such
    that they perform well on real data~\cite{general_biosig_gans_2023}.
In addition to this, it was noted that the generative models were readily able to capture the noise
    of the input data~\cite{general_biosig_gans_2023}.

It was found that although GANs have found lots of use traditionally, the number of papers in medical
    imaging that utilise VAEs and DMs has increased in recent years.
For DMs in particular, there has been a substantial increase in papers, which authors attributed to
    their ability to generate high-quality images with good mode
    coverage~\cite{medical_imaging_review_2023}.
Despite the abundance of diffusion models in medical imaging, we could not find, to the best of our knowledge, any use in biomedical audio signals, leaving room for exploration.

\subsection{Conditional Denoising Diffusion Probabilistic Models}

Denoising Diffusion Probabilistic models (DDPM) are a type of diffusion model that follows a Markov
    process that continuously noises the input, with the network learning to reverse this process by
    estimating the noise that was added.
{Further details on the mechanisms of these models are found in \ref{ap:diff}.}
Conditional diffusion models for conditional audio generation can be adapted 
    from the diffusion model setup in \cite{ho2020denoising}.

\subsubsection{WaveGrad}

\begin{figure*}[htbp]
    \begin{subfigure}{0.33333\textwidth}
        \centering
        \includegraphics[width=\linewidth]{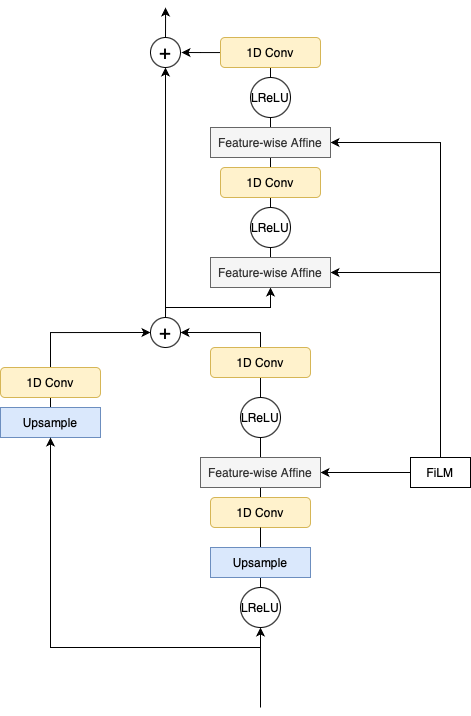}
        \caption{UBlock Module Architecture.}\label{fig:ublock}
    \end{subfigure}%
    \begin{subfigure}{0.33333\textwidth}
        \centering
        \includegraphics[width=0.8\linewidth]{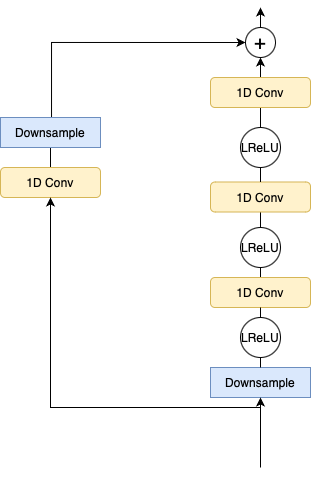}
        \caption{DBlock Module Architecture.}\label{fig:dblock}
    \end{subfigure}%
    \begin{subfigure}{0.33333\textwidth}
        \centering
        \includegraphics[width=0.8\linewidth]{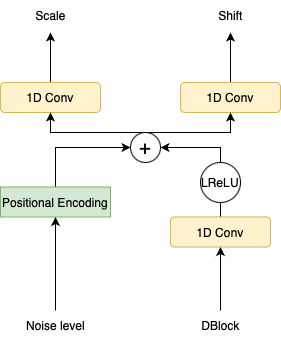}
        \caption{FiLM Module Architecture.}\label{fig:film}
    \end{subfigure}%

    \caption{WaveGrad Module Architectures.}\label{fig:modules}
\end{figure*}

\begin{figure}[H]
    \centering
    \includegraphics[width=.4\linewidth]{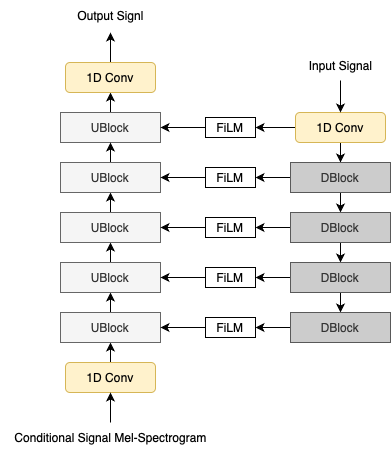}
    \caption{WaveGrad Architecture.}\label{fig:wavegrad-arch}
\end{figure}

WaveGrad is a DDPM for audio synthesis using conditional generation.
The model utilises the architecture consisting of multiple upsampling blocks (UBlocks) and
    downsampling blocks (DBlocks), with the input signal and the conditioning signal as inputs into
    the network.
The conditioning signal is converted to a mel-spectrogram representation before being input to 
    the model~\cite{wavegrad}.
These UBlocks and DBlocks follow the architecture of the upsampling and downsampling blocks utilised in the Generative Adversarial Network text-to-speech (GAN-TTS) model \cite{gan-tts}.
The feature-wise linear modulation (FilM) modules combine information from the noisy waveform and the conditioning mel-spectrogram \cite{wavegrad}.
The UBlock, DBlock and feature-wise linear modulation (FiLM) modules are shown in
    \Cref{fig:modules}, with \Cref{fig:wavegrad-arch} showing the entire WaveGrad architecture.
The loss function is based on the difference between the noise added in each step of the forward
    diffusion process and the noise predicted during the reverse process~\cite{wavegrad} as
    described in \Cref{eqn:wave-loss}, with the Markov process being conditioned on the continuous
    noise level instead of the time-step.
Also, note that the L1 norm was used over the L2 norm as it was found to provide better training
    stability~\cite{wavegrad}.
WaveGrad only includes a local conditioner in the form of a conditioning signal.

\begin{equation}
    \label{eqn:wave-loss}
    \E_{\overline{\alpha}, \epsilon}\left[ \norm{\epsilon_{\theta}\left(\sqrt{\overline{\alpha}}\bold{y_0}
        + \sqrt{1-\overline{\alpha}}\epsilon , \bold{x}, \sqrt{\overline{\alpha}}\right) - \epsilon_t}_1
        \right]
\end{equation}

\subsubsection{DiffWave}

\begin{figure}[H]
    \centering
    \hspace*{-1.5cm}
    \includegraphics[width=0.8\linewidth]{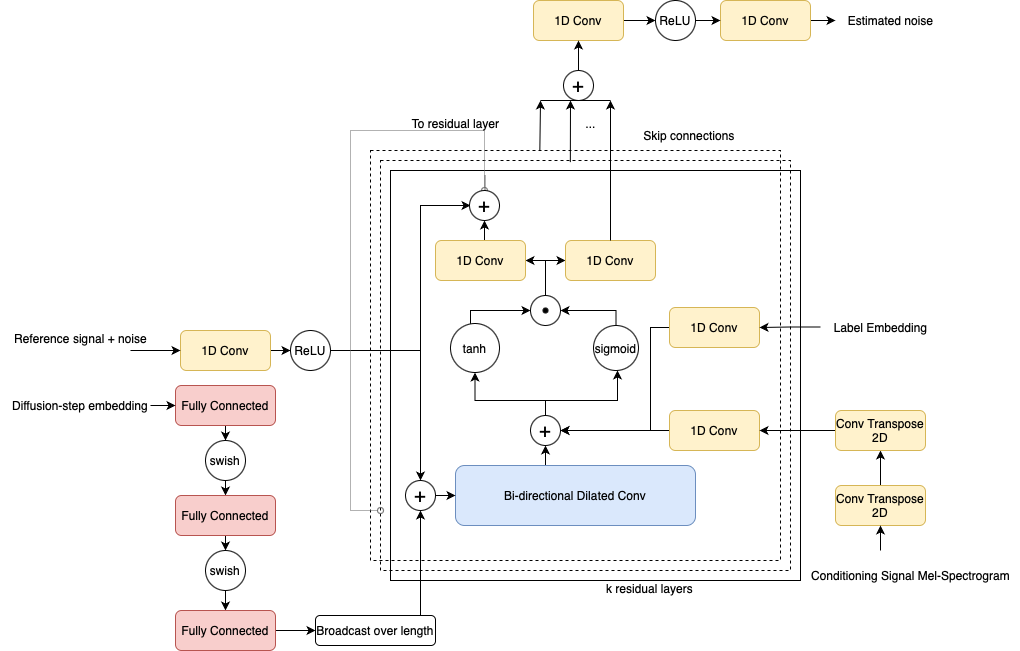}
    \caption{DiffWave Architecture.}\label{fig:diffwaveArch}
\end{figure}

DiffWave is another DDPM for raw audio synthesis with conditional and unconditional generation.
The loss function utilises a single ELBO-based training objective without auxiliary
    losses~\cite{diffwave}, as described in \Cref{eqn:diff-loss}.
One-dimensional convolutions are used on the input and conditioning signals that go through multiple
    fully connected layers.
The model contains a WaveNet~\cite{wavenet} \textit{backbone}, consisting of bi-directional dilated convolutions and residual layers and
    connections.
The architecture is shown in \Cref{fig:diffwaveArch}.
DiffWave can be used for both conditional and unconditional generation.
For conditional generation, it uses a local conditioning signal and a global conditioner (discrete
    labels)~\cite{diffwave}.

\begin{equation}
    \label{eqn:diff-loss}
    \E_{t, \epsilon}\left[ \norm{\epsilon_{\theta}\left(\sqrt{\overline{\alpha}_t}\bold{y_0} +
        \sqrt{1-\overline{\alpha}_t}\epsilon , \bold{x}, t\right) - \epsilon_t}_1 \right]
\end{equation}

\section{Materials and Methods}

To achieve a more robust model, the augmented training dataset must first be created. 
    \Cref{fig:DAArchitecture} depicts the dataset creation process.
Once this dataset is created, various classification models can be trained and evaluated to measure
    the increase in ID and OOD performance.

\begin{figure}[htbp]
    \centering
    \begin{tikzpicture}[every node/.style={text width=2.5cm, align=center},auto,node distance=1.5cm]

\usetikzlibrary{positioning, arrows.meta, decorations.markings}

\tikzstyle{vecArrow} = [thick, decoration={markings,mark=at position
   1 with {\arrow[semithick]{open triangle 60}}},
   double distance=1.4pt, shorten >= 5.5pt,
   preaction = {decorate},
   postaction = {draw,line width=1.4pt, white,shorten >= 4.5pt}]
\tikzstyle{innerWhite} = [semithick, white,line width=1.4pt, shorten >= 4.5pt]

\tikzset{%
    block/.style= {rectangle, draw, fill=white!20, minimum height=2em, minimum width=4em},
    arrow/.style= {thick, ->, >=stealth},
    line/.style= {thick, -},
}

\node [block] (pcg) {Original Training Dataset};
\node [block, below left=0.5cm and 0.5cm of pcg.south] (pre) {Preprocessing};
\node [block, below of=pre] (seg) {Heart Sound Segmentation};
\node [block, below =2cm of pre.south] (cce) {Cardiac Cycle Extraction};
\node [block, below of=cce] (glm) {Generative Learning Model};
\node [block, right=1cm of pre.east] (ta) {Traditional Augmentation};
\node [block, below right=0.5cm and 0.5cm of glm.south] (ad) {Augmented Training Dataset};

\draw [vecArrow] (pcg) -| (ta);
\draw [vecArrow] (pcg) -| (pre);
\draw [vecArrow] (pre) -- (seg);

\draw [vecArrow] (seg) -- (cce);

\draw [vecArrow] (seg) -- (cce);

\draw [vecArrow] (cce) -- (glm);
\draw [vecArrow] (glm) |- (ad);
\draw [vecArrow] (ta) |- (ad);

\end{tikzpicture}
    \caption{Data Augmentation Architecture}\label{fig:DAArchitecture}
\end{figure}

\subsection{Datasets}

\subsubsection{PhysioNet and Computing in Cardiology Challenge 2016 Dataset}

The PhysioNet and Computing in Cardiology Challenge 2016 (CinC) was an international competition that 
    aimed to encourage the development of heart sound classification algorithms~\cite{cinc_nov16}.
The data was sourced from nine independent databases but excluded a database focused on
    fetal and maternal heart sounds~\cite{cinc_nov16}.
Across the nine databases, there are \SI{2435}{} recordings sourced from \SI{1297}
    patients~\cite{cinc_nov16}.
Excluding the aforementioned database and splitting longer recordings into smaller samples, there
    were in total \SI{4430}{} samples from \SI{1072} patients, equating to \SI{233512}{} heart
    sounds, \SI{116865}{} heart beats, and nearly \SI{30}{} hours of recordings used in the
    competition~\cite{cinc_mar16}.
At the time of (their) publication, this amounted to the largest open-access heart sound database in
    the world~\cite{cinc_mar16}.

The recordings were resampled to \SI{2000}{\hertz} for the competition and only one PCG lead was
    used, with the exception of training-set \textit{a}, which includes ECG~\cite{cinc_mar16}.

\begin{table}[htbp]
    \caption{Summary of Challenge Data}\label{tab:DataSummaryTable}
    \centering
    \scalebox{\resultsscale}
    {\begin{tabular}{cllrrr}
    \toprule
    \multicolumn{3}{c}{\textbf{Database Information}} & \multicolumn{3}{c}{\textbf{Proportion of Recordings (\%)}} \\
    Challenge Use & Dataset & Source Database & Abnormal & Normal & Unsure \\
    \midrule
    \multirow{6}{*}{\textit{Training}} & training-a & MITHSDB  & 67.5 & 28.4 & 4.2  \\
                                       & training-b & AADHSDB  & 14.9 & 60.2 & 24.9 \\
                                       & training-c & AUTHHSDB & 64.5 & 22.6 & 12.9 \\
                                       & training-d & UHAHSDB  & 47.3 & 47.3 & 5.5  \\
                                       & training-e & DLUTHSDB & 7.1  & 86.7 & 6.2  \\
                                       & training-f & SUAHSDB  & 27.2 & 68.4 & 4.4  \\
                                       &            & Average  & 18.1 & 73.0 & 8.8 \\
                                       \midrule
    \multirow{6}{*}{\textit{Test}}    & test-b & AADHSDB  & 15.6 & 48.8 & 35.6 \\
                                      & test-c & AUTHHSDB & 64.3 & 28.6 & 7.1  \\
                                      & test-d & UHAHSDB  & 45.8 & 45.8 & 8.3  \\
                                      & test-e & DLUTHSDB & 6.7  & 86.4 & 6.9  \\
                                      & test-g & TUTHSDB  & 18.1 & 81.9 & 0.0    \\
                                      & test-i & SSHHSDB  & 60   & 34.3 & 5.7  \\
                                      &        & Average  & 12.0 & 77.1 & 10.9 \\
                                      \bottomrule
\end{tabular}
}
\end{table}

Recordings were divided into either 
    \textit{normal} (healthy),
    \textit{abnormal} (diagnosed with CVD or other cardiac problems),
    or \textit{unsure} (low quality signals)~\cite{cinc_nov16}.
A summary of the data, shown in \Cref{tab:DataSummaryTable},
    was adapted from~\cite{cinc_mar16} and~\cite{cinc_nov16}.
These datasets also include additional information, such as individual disease diagnoses and
    annotations of the heart cycles.
These can be used to assist with the data augmentation.

\subsubsection{Synchronised Multichannel PCG and ECG dataset}

Recently, synchronised multichannel PCG and ECG (SMPECG) data has been collected from an EPCG device
    that consists of seven PCG and one lead-I ECG sensors~\cite{RongFynnNordholm2023PreScreeningCAD}.
Using this device, data was collected from 105 subjects, of which 46 were diagnosed with coronary
    artery disease.
Ten seconds of audio were recorded for each subject, during which the subjects were instructed not
    to breathe to eliminate lung sounds from the recording.
This data was collected in a clinical environment with background noise and non-optimal sensor placement as it is designed for ease of use, making it a
    challenging dataset for classification, {which is representative of a real-world dataset}.
As only single channel PCG is available in the other datasets, only a single channel (channel 2) was
    used for this dataset.

\subsubsection{Incentia Dataset}

Along with the training-a dataset used for the inputs for training the generative models, the
    incentive dataset~\cite{icentia} was utilised to provide unique unseen ECG to generate an
    accompanying PCG signal.
This data set contains 11,000 patients and 2,774,054,987 labelled heartbeats at a sample rate of
    \SI{250}{\hertz} with 541,794 segments.
Each beat was classified with a type from normal, premature atrial contraction, premature
    ventricular contraction and rhythm from normal sinusal rhythm, atrial fibrillation and atrial
    flutter.

\subsubsection{Further Datasets}

To improve the model's robustness against noise, one of the stages of augmentation introduces noise
    from other PCG and ECG datasets.
These are the electro-phono-cardiogram (EPHNOGRAM) dataset~\cite{ephnogram} for PCG and the  Massachusetts Institute of Technology - Beth Israel Hospital (MIT-BIH) dataset~\cite{mit} for ECG\@.
The EPHNOGRAM dataset comprises 24 healthy adults and contains recordings taken during stress tests
    and at rest~\cite{ephnogram}.
The MIT-BIH dataset contains 12 half-hour ECG recordings and three half-hour recordings of noise
    typical in ambulatory ECG recordings, where this noise is used for augmentation~\cite{mit}.

\subsection{Signal Augmentation}

The augmentation procedure of the PCG and ECG signals is shown in \Cref{fig:epcg_augment}.
The time stretching augmentation is synchronised to ensure that they are both stretched the same
    amount, with the black lines representing the flow of the ECG data and the white lines
    representing the flow of PCG data.
Augmentation stages have different percentage chances of occurring, where the chances chosen were determined to provide 
    the widest variety of augmented signals after every stage has been completed whilst also resulting in the best
    performance.
The augmentations vary slightly between PCG and ECG to best meet the physiological constraints.

\begin{figure}[htbp]
    \centering
    \begin{tikzpicture}[every node/.style={text width=2.5cm, align=center},auto,node distance=1.5cm]

\usetikzlibrary{positioning, arrows.meta, decorations.markings}

\tikzstyle{vecArrow1} = [thick, decoration={markings,mark=at position
   1 with {\arrow[semithick]{open triangle 60}}},
   double distance=1.4pt, shorten >= 5.5pt,
   preaction = {decorate},
   postaction = {draw,line width=1.4pt, white,shorten >= 4.5pt}]
\tikzstyle{innerWhite} = [semithick, white,line width=1.4pt, shorten >= 4.5pt]
\tikzstyle{vecArrow2} = [thick, decoration={markings,mark=at position 1 with {\arrow[semithick]{triangle 60}}}, double distance=1.4pt, shorten >= 5.5pt, preaction={decorate}, postaction={draw,line width=1.4pt, black,shorten >= 4.5pt}]

\tikzset{%
    block/.style= {rectangle, draw, fill=white!20, minimum height=2em, minimum width=4em},%
    arrow/.style= {thick, ->, >=stealth},%
    line/.style= {thick, -},%
}

\node [block] (pcg) {Original PCG Data};
\node [block, right=1cm of pcg.east] (ecg) {Original ECG Data};
\node [block, below of=pcg] (hpss) {HPSS Emphasis Filter};
\node [block, below of=hpss] (noise1p) {White Noise};
\node [block, below of=ecg] (noise1e) {White Noise};
\node [block, below of=noise1e] (wander) {Baseline Wander};
\node [block, below left=0.5cm and 0.5cm of wander.south] (stretch) {Time Stretching};
\node [block, below left=0.5cm and 0.5cm of stretch.south] (am) {Amplitude Modulation};
\node [block, below of=am] (noise2p) {White Noise};
\node [block, right=1cm of noise2p.east] (noise2e) {White Noise};
\node [block, below of=noise2p] (peqp) {Parametric Equalisation band emphasis};
\node [block, below of=noise2e] (peqe) {Parametric Equalisation band emphasis};
\node [block, below of=peqp] (clinicalp) {Clinical Noise};
\node [block, below of=peqe] (clinicale) {Clinical Noise};
\node [block, below of=clinicalp] (augmentp) {Augmented PCG Data};
\node [block, below of=clinicale] (augmente) {Augmented ECG Data};

\draw [vecArrow1] (pcg) -- (hpss);
\draw [vecArrow1] (hpss) -- (noise1p);
\draw [vecArrow2] (ecg) -- (noise1e);
\draw [vecArrow1] (noise1p) |- (stretch);
\draw [vecArrow2] (noise1e) -- (wander);
\draw [vecArrow2] (wander) |- (stretch);
\draw [vecArrow1] (stretch) |- (am);
\draw [vecArrow2] (stretch) |- (noise2e);
\draw [vecArrow1] (am) -- (noise2p);
\draw [vecArrow1] (noise2p) -- (peqp);
\draw [vecArrow2] (noise2e) -- (peqe);
\draw [vecArrow1] (peqp) -- (clinicalp);
\draw [vecArrow2] (peqe) -- (clinicale);
\draw [vecArrow1] (clinicalp) -- (augmentp);
\draw [vecArrow2] (clinicale) -- (augmente);

\end{tikzpicture}
    \caption{PCG and ECG traditional augmentation procedure}\label{fig:epcg_augment}
\end{figure}

The PCG signals are augmented in various ways: harmonic percussive source separation (HPSS) for
    emphasis on certain parts of the signal, time stretching, emphasis on certain bands of the
    signal using a parametric equalisation (EQ) filter and introducing noise from the EPHNOGRAM
    dataset~\cite{ephnogram}.
Before these operations are applied, the signals are normalised to have a zero mean and be between
    -1 and 1.
Shown in \Cref{fig:epcg_augment} is the augmentation procedure applied to PCG data, noted with the
    white lines.

The HPSS  has a 75\% chance of occurring and works by extracting harmonic and percussive components
    of the signal with varying thresholds to extract different parts of the signal.
    The HPSS implementation is from the librosa v0.1.0 Python library~\cite{librosa, hpss}.
$\bold{X}(t,k)$ denotes the short-time Fourier transform (STFT) of the signal $\bold{x}(t)$, defined as

\begin{equation}
    \bold{X}(t, k) = \sum_{n=0}^{N-1} \bold{w}(n) \bold{x}(n + tH) \exp\left(-2 \pi j k n / N\right)
\end{equation}

\noindent where $\bold{w}$ is a sine-window, $H$ represents the hop size and $N$ is the window length and the length of
    the discrete Fourier transform.

Firstly, the STFT of the signal is calculated, with the parameters chosen randomly from a window
    length of 512, 1024 and 2048 with equal probability.
A hop length was randomly chosen from 16, 32, 64, and 128 with uniform distribution.

Following this, the harmonic and percussive components are then extracted from the following,

\begin{equation}
    \tilde{\bold{Y}}_h(t, k) = median(\bold{X}(t-\ell_h, k),\ldots{},\bold{X}(t+\ell_h,k))
\end{equation}

\begin{equation}
    \tilde{\bold{Y}}_p(t, k) = median(\bold{X}(t, k- \ell_p),\ldots{},\bold{X}(t,k+ \ell_p))
\end{equation}

\begin{equation}
    \bold{M}_h(t,k) = 
    \begin{cases} 
    1, & \text{if } \frac{\tilde{\bold{Y}}_h(t,k)}{\tilde{\bold{Y}}_p(t,k) + \eta} > \lambda_h \\
    0, & \text{otherwise}
    \end{cases}
\end{equation}

\begin{equation}
    \bold{M}_p(t,k) = 
    \begin{cases} 
    1, & \text{if } \frac{\tilde{\bold{Y}}_p(t,k)}{\tilde{\bold{Y}}_h(t,k) + \eta} \ge \lambda_p \\
    0, & \text{otherwise}
    \end{cases}
\end{equation}

\begin{equation}
    \bold{X}_h(t,k) = \bold{X}(t,k) \cdot \bold{M}_h(t,k)
\end{equation}

\begin{equation}
    \bold{X}_p(t,k) = \bold{X}(t,k) \cdot \bold{M}_p(t,k)
\end{equation}

\noindent where $\bold{X}_h(t,k)$ is the harmonic component, $\bold{X}_p(t,k)$ is the percussive component $\eta$ is a
    small number added to avoid a divide by 0 error~\cite{hpss}.
$\bold{x}_h(t)$ and $\bold{x}_p(t)$ are the inverse STFT (ISTFT) of $\bold{X}_h(t,k)$ and $\bold{X}_p(t,k)$.
If the thresholds, $\lambda_h > 1$ or $\lambda_p > 1$, there will be some part of the spectrum that is not a harmonic or percussive component of the signal but a residual component that appears as textured noise. As the abnormalities to be detected are from diseases that produce more percussive or harmonic sounds, these residuals can be ignored without important information loss that would negatively impact the ability of a classifier to classify these sounds.

The first set have parameters $\lambda_{h} = rand(1, 2)$, $\lambda_{p} = rand(1, 2)$, $\ell_{h} =
    randint(5, 30)$, and $\ell_{p} = randint(5, 30)$.
$rand$ denotes a random floating point number chosen uniformly between the two bounds, and $randint$
    is an integer uniformly chosen between those bounds.
The second set are then extracted from $\bold{X}_{h}(t,k)$ and $\bold{X}_{p}(t,k)$.
$\bold{X}_{hh}(t,k)$ and
    $\bold{X}_{hp}(t,k)$ are the harmonic and percussive components of $\bold{X}_{h}(t,k)$ and $\bold{X}_{ph}(t,k)$ and $\bold{X}_{pp}(t,k)$ the harmonic and percussive components of $\bold{X}_{p}(t,k)$.
The second stage of decomposition uses parameters of $\lambda_{hh} = rand(1, 4)$, $\lambda_{hp} = rand(1,
    4)$, $\lambda_{ph} = rand(1, 4)$, $\lambda_{pp} = rand(1, 4)$, $\ell_{hh} = randint(5, 30)$,
    $\ell_{hp} = randint(5, 30)$, and $\ell_{ph} = randint(5, 30)$, $\ell_{pp} = randint(5, 30)$.

The ISTFT is then applied to each component before reconstructing the signal as,

\begin{equation}
    \label{eqn:hpss}
    \begin{split}
        \bold{s}_{HPSS}(t) = a_{hh}\bold{x}_{hh}(t) + a_{hp}\bold{x}_{hp}(t) +
            a_{ph}\bold{x}_{ph}(t) + a_{pp}\bold{x}_{pp}(t)
    \end{split}
    \end{equation}

\noindent where $a_{hh} = rand(0.01, 10)$, $a_{hp} = rand(0.01, 10)$,
    $a_{ph} = rand(0.01, 10)$, $a_{pp} = rand(0.01, 10)$.

This two stage decomposition and reconstruction described in \Cref{eqn:hpss} is done twice to create $\bold{s}_{HPSS_1}(t)$ and $\bold{s}_{HPSS_2}(t)$, which are then combined to get the final augmented signal
    $\bold{s}_{HPSS_{final}}(t)$,

\begin{equation}
    \bold{s}_{HPSS_{final}}(t) = \bold{s}_{HPSS_1}(t) + a_{HPSS}\bold{s}_{HPSS_2}(t)
\end{equation}

\noindent where $a_{HPSS} = rand(0.01, 0.05)$.
The use of these parameters was determined by inspection to ensure the signals remain realistic.

Next, there is a 7.5\% chance of introducing noise to the signal, as defined in the equation below,
    where $\bold{s}_{HPSS}(t)$ is the signal after the HPSS augmentation stage, $\bold{s}_{SN}(t)$ is the augmented signal and 
    $\bold{r}(t) \sim \mathcal{N}(\mu, \sigma I)$, $\sigma = rand\_choice(0.01, 0.001, 0.0001)$ and $\mu =
    rand(0, 0.1)$.
    Note that $\bold{s}_{HPSS}(t)$ may not have had the HPSS augmentation applied as it depends on the random chance.
    $rand\_choice()$ denotes a random choice from those numbers with equal probability.
\begin{equation}
    \label{eqn:noise}
    \bold{s}_{SN}(t) = \bold{s}_{HPSS}(t) + \bold{r}(t)
\end{equation}

Following this, there is a 25\% chance of adding in a time warp.
    This time warp will stretch the signal randomly to either 1.004 times the length or 1.006 times
    the length of the original signal.
It is noted that a time warp with the same factor will be applied to both the PCG and ECG\@.

There is then a 75\% chance of adding in amplitude modulation.
    The modulation is done as described in \Cref{eqn:modulation}, where $b_{AM_1} = rand(0.01,
    0.25)$, $b_{AM_2} = rand(0.01, 0.25)$, $c_{AM_1} = rand(0.05, 0.5)$, $c_{AM_2} = rand(0.001,
    0.05)$, $d_{AM_1} = rand(0,1)$, $d_{AM_2} = rand(0,1)$ and $s_{TS}(t)$ is signal after the time stretch augmentation stage, which depending on the random chance may have been time-stretched.
\begin{equation}
    \label{eqn:modulation}
    \bold{s}_{AM} = \bold{s}_{TS}(t) \cdot \left(1 + b_{AM_1} \sin\left(2\pi c_{AM_1} t + d_{AM_1}\right) + b_{AM_2}
        \sin\left(2 \pi c_{AM_2} t + d_{AM_2}\right)\right)
\end{equation}

Next, there is another 7.5\% chance of introducing the same noise as done in \Cref{eqn:noise}.
Following this, there is a 25\% chance of applying 
    parametric equalisation to boost frequency bands.
Given the frequency range of \SIrange{2}{500}{\hertz},
    the bandwidth is randomly selected between 5\% and 20\% of this range,
    and the signal is attenuated using a bandpass filter.
After repeating this process 5 times,
    the filtered signal and original signal
    are summed and normalised.

Lastly, real noise from the EPHNOGRAM dataset is introduced.
The introduced noise from the EPHNOGRAM is clinical noise extracted from some of the recordings in
    this dataset.
This augmentation occurs 50\% of the time.

The ECG signals are also augmented in numerous ways; these include introducing random noise, adding
    baseline wander, time stretching, adding noise from the MIT-BIH dataset, and emphasising certain
    signal bands. \Cref{fig:epcg_augment} shows the order of processing on the ECG, indicated with
    the black lines.

Random noise is applied the same way as the PCG noise, as defined in \Cref{eqn:noise}, with this
    augmentation occurring with a probability of 7.5\%.
Next, a baseline wander is added 30\% of this time.
    This is done as described in \Cref{eqn:wander}, where $b_{BW_1} = rand(0.01, 0.2)$, $b_{BW_2} =
    rand(0.01, 0.2)$, $c_{BW_1} = rand(0.05, 0.5)$, $c_{BW_2} = rand(0.001, 0.05)$, $d_{BW_1} =
    rand(0,1)$, $d_{BW_2} = rand(0,1)$. $\bold{s}_{SN_E}(t)$ is the ECG signal after the random noise augmentation stage, which may include the random noise as per the random chance.
\begin{equation}
    \label{eqn:wander}
    \bold{s}_{BW}(t) = \bold{s}_{SN_E}(t) + b_{BW_1} \sin\left(2\pi c_{BW_1} t + d_{BW_1}\right) + b_{BW_2}\sin\left(2\pi
        c_{BW_2} t + d_{BW_2}\right)
\end{equation}

%PUT SOMETHING SHOWING THIS AUGMENTATION MAYBE?}

Following this, there is a 25\% chance of a timewarp between 1 and 1.06 times the original
    signal.
    It is noted that a timewarp with the same factor will be applied to both the PCG and ECG\@.
Then, the same parametric equalisation, as with the PCG, is applied between \SI{0.25}{\hertz} and
    \SI{100}{\hertz}.

%PUT SOMETHING SHOWING THIS AUGMENTATION MAYBE?}

Lastly, noise from the MIT-BIH database is added, occuring 50\% of the time.
This is noise from the ECG sensors taken from recordings in the MIT-BIH database.

%PUT SOMETHING SHOWING THIS AUGMENTATION MAYBE?v}

\subsection{Synthetic Audio Generation}

Synthetic signals were generated using the mel-spectrogram of the ECG signal as a conditioner for both the     
    WaveGrad~\cite{wavegrad} and DiffWave~\cite{diffwave} diffusion models.
They are trained before data is generated for use.
These diffusion models generated data for 3200 patients, 800 abnormal and 2400 normal, with three segments
    used to train the classification models.
This is done to reduce the effect of overfitting to the synthetic signals.
The ECG signals for conditioning were taken from the icentia database~\cite{icentia} to introduce
    new data, with abnormal ECG used for abnormal PCG\@.
The generative models were trained to create individual conditions and make them more realistic
    using additional labels from the dataset.
To get around the lack of training data, the order of heart cycles was rearranged to increase
    training diversity.
DiffWave and WaveGrad models were trained on an Nvidia RTX 4090 for 24 hours.
The parameters for the DiffWave model that differ from the default are shown below in
    \Cref{tab:diffwave_param}.
Parameters used for the WaveGrad model that differ from the default are shown in Table
    \Cref{tab:wavegrad_param}.
Both models differ slightly from their base implementations as they use a custom global conditioner.
Additional global conditioners were added for specific abnormalities or lack of abnormalities, such
    as mitral valve prolapse, innocent or benign murmurs, aortic disease, miscellaneous conditions,
    and normal.

\begin{table}[htbp]
    \caption{\label{tab:diffwave_param}DiffWave parameters}
    \centering
    \begin{tabular}{lr}
        \toprule
        \textbf{Parameter} & \textbf{Value}\\[0pt]
        \midrule
        Residual layers & 30\\[0pt]
        Residual channels & 64\\[0pt]
        Dilation cycle length & 10\\[0pt]
        Embedding dimension & 32\\[0pt]
        Batch size & 8\\[0pt]
        Learning rate & 2e-4\\[0pt]
        Noise schedule & T=50, linearly spaced [1e-4, 5e-2]\\[0pt]
        Inference noise schedule & \{1e-4, 1e-3, 1e-2, 5e-2, 2e-1, 5e-1\}\\[0pt]
        \bottomrule
    \end{tabular}
\end{table}

\begin{table}[htbp]
    \caption{\label{tab:wavegrad_param}WaveGrad parameters}
    \centering
    \begin{tabular}{lr}
        \toprule
        \textbf{Parameter} & \textbf{Value}\\[0pt]
        \midrule
        Embedding dimension & 32\\[0pt]
        Batch size & 8\\[0pt]
        Learning rate & 2e-4\\[0pt]
        Noise schedule & T=1000, linearly spaced [1e-6, 1e-2]\\[0pt]
        \bottomrule
    \end{tabular}
\end{table}

To ensure a diversity of training examples, various heart cycles were occasionally
    rearranged for each patient for each minibatch during training.
This was done inside a custom collator, with a 75\% chance of rearranging the heart cycles.
Heart cycles could be rearranged in three ways with equal probability.
The first will take groupings of many cycles and then randomly rearrange these large groups.
These first groups would have a size of half of the total number of heart cycles within that signal.
Secondly, groupings of 1 to 4 heart cycles were chosen randomly and used to rearrange the signal.
Finally, the third way involved rearranging each heart cycle.

Although this rearranging can violate physiological constraints, it was found that this helped the model learn a better representation of the data and improved classification results when trained on the synthetic data. 

The signals were then bandpass filtered between \SI{2}{\hertz} to \SI{500}{\hertz} for PCG and
    \SI{0.25}{\hertz} to \SI{100}{\hertz} for ECG, the conditioning signal.
A mel-spectrogram of the ECG was created as the local conditioning signal.
The mel-spectrogram was created using a sample rate of \SI{4}{\kilo\hertz}, window length 1024, hop
    length 256, and 80 mel bins.
Crossfading was used to ensure minimal audio artifacts when rearranging heart cycles.
As the signals are joined when they are both in the same state, the end of the cycle in the
    diastole phase, they are assumed to be roughly correlated.
The crossfade occurs between the last 40 samples of the first signal, $ -1 \le t \le 0$, and the
    first 40 samples from the second signal, $0 \le t \le 1$.
If one of the signals has a low variance, then a simple linear crossfade is used between the two.
A linear crossfade can be described from \Cref{eqn:linear_fade,eqn:fade} below,

\begin{equation}
    \label{eqn:linear_fade}
    \bold{f}(t) = 1/2 + t/2,\;\;-1 < t < 1
\end{equation}

\begin{equation}
    \label{eqn:fade}
    \bold{v}(t) = \bold{f}(t)\bold{y}(t) + \bold{f}(-t)\bold{x}(t)
\end{equation}

\noindent where $f$ is the crossfade function, $v$ is the final spliced signal, $x$ is the last 40 samples
    from the first signal, and $y$ is the first 40 samples from the second signal.

\begin{figure}
    [h] \centering \includegraphics[width=\linewidth]{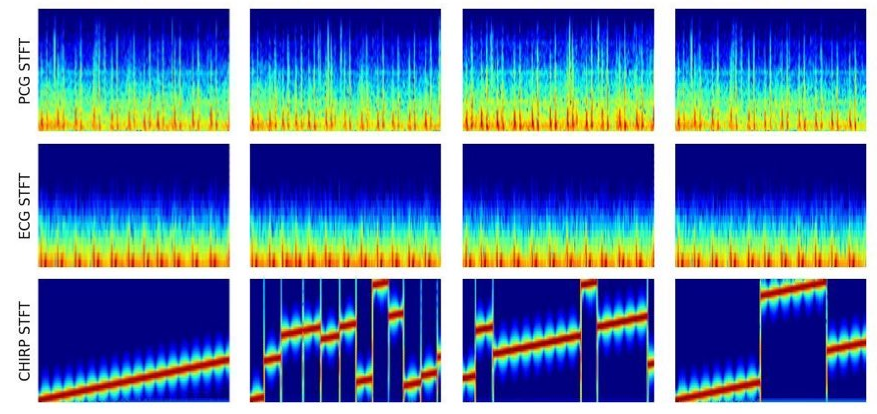} \caption{Rearranged
        heart cycles with crossfade.}\label{fig:rearrange}
\end{figure}

Otherwise, the following crossfade function will be used to ensure a crossfade is applied that
    represents how correlated the two signals are.
For two fully uncorrelated signals, a constant power crossfade would be desired, and for two fully
    correlated signals, a constant voltage crossfade would be desired and something in between if
    not fully correlated or uncorrelated.
Assuming that the crossfade function is deterministic, the two signals are a random
    process.
Along with the assumption, the mean power of the signals at the point of crossfading is equal as
    they are being crossfaded when in the same phase of the heart cycle.
This allows the following generalised crossfade function~\cite{crossfade} to be used to satisfy a
    crossfade related to the signals' correlation.
The crossfade is defined in \Cref{eqn:hann_fade,eqn:odd_fade,eqn:general_fade},

\begin{equation}
    \label{eqn:hann_fade}
    \bold{o}(t) = \frac{9}{16}\sin\left(\frac{\pi}{2}t\right) +
        \frac{1}{16}\sin\left(\frac{3\pi}{2}t\right),\;\;\; -1 < t < 1
\end{equation}

\begin{equation}
    \label{eqn:odd_fade}
    \bold{e}(t) = \sqrt{\frac{1}{2(1+r)} - \left(\frac{1-r}{1+r}\right)\bold{o}{(t)}^2}
\end{equation}

\begin{equation}
    \label{eqn:general_fade}
    \bold{f}(t) = \bold{o}(t) + \bold{e}(t)
\end{equation}

\noindent where $e$ is the even component of the crossfade function, and $o$ is the odd component, and $r$ is
    the correlation coefficient of the two signals at zero lag and $0 \le r \le 1$.
The crossfade is then interpolated to double the length using a univariate spline, with a degree of 3 and a smoothing factor equal to the length of the signal.
The implementation is the scipy implementation of the univariate spline \cite{scipy}.
The final signal consists of the first signal before the last 40 samples, the crossfaded and
    interpolated signal, and the second signal after the first 40 samples.  \Cref{fig:rearrange}
    demonstrates the effect that this crossfade has on reducing artifacts.
Rearranging of the heart cycles can be seen through the rearranging of the chirp in the last
    row.
The first column shows the original signal, the second shows the rearranging of all heart cycles, the
    third shows the rearranging of a few heart cycles, and the final shows the rearranging of larger
    groups of heart cycles.

\subsection{Classification Model}

The model used to test the augmented dataset is a convolutional neural network-based model finetuned from ResNet trained on
    ImageNet~\cite{milan}.
The purpose of choosing this model is not to show its better performance in classification but to demonstrate the capability of the proposed data augmentation methods.
Before the signals are passed into the convolutional neural network (CNN), the PCG signal is bandpass filtered between
    \SI{45}{\hertz} and \SI{400}{\hertz}.
The ECG signal is bandpass filtered between 25Hz and 100Hz.
The signals also then undergo normalisation.
A spectrogram is created from the signal before being passed to the model, with a window length of 100 and a hop length of 50.
This spectrogram is created based on \SI{1.5}{\second} of audio, with each being referred to as a
    fragment, with the training objective to maximise accuracy on the fragment level.
From the synthetic data, only three fragments of \SI{1.5}{\second} audio are taken to ensure reduced
    overfitting to the synthetic data.
These \SI{1.5}{\second} fragments differ from the original model~\cite{milan} which took in a single
    heart cycle.
This change has been done to reduce the need for accurate segmentation.
For testing the subject level, the outputs from the classification are averaged between all
    fragments before the classification is made, as was done previously. The Adam optimiser is used for training along with a cyclic triangular learning rate scheduler with
    parameters below in \Cref{tab:optim_params}.

\begin{table}[htbp]
    \caption{\label{tab:optim_params} Adam Optimiser Parameters}
    \centering
    \begin{tabular}{lr}
        \toprule
        \textbf{Parameter} & \textbf{Value} \\[0pt]
        \midrule
        initial learning rate & $0.001$\\
        betas & $(0.9, 0.999)$\\
        epsilon & $10^{-8}$ \\
        weight decay & $10^{-3}$\\
        learning rate step size up & $2$ \\
        learning rate step size down & $2$ \\
        max learning rate & $10^{-3}$ \\
        \bottomrule
    \end{tabular}
\end{table}

During the model's training on the original dataset, as a CNN is being finetuned, only 10 epochs are
    used in which the best weights are chosen from the highest MCC value from the validation set to
    reduce overfitting.
The model is only updated for each dataset if it performed better on the validation set than
    previously.
A schedule is used to reduce the overfitting of the synthetic data for training on the augmented
    dataset.
This schedule can be found below in \Cref{tab:training_schedule} and was experimentally determined
    to provide the best results, where max-mix is all of the data with no augmentations being applied to the original dataset and 3
    augmentations applied to the DiffWave and WaveGrad data.
From the synthetic data, only three random segments were taken to ensure the model does not overfit
    to the synthetic data.
The max-aug data is the original data with 30 augmentations being applied and no synthetic data.

\begin{table}[htbp]
    \caption{\label{tab:training_schedule} Training Schedule}
    \centering
    \begin{tabular}{lr}
        \toprule
        \textbf{Data} & \textbf{Epochs}\\[0pt]
        \midrule
        max-mix & 8 \\
        max-aug & 8 \\
        max-mix & 8 \\
        max-aug & 8 \\ 
        max-mix & 8 \\
        max-aug & 8 \\ 
        max-mix & 16 \\
        max-aug & 16 \\
        max-mix & 16 \\
        max-aug & 16 \\ 
        max-mix & 16 \\
        max-aug & 16 \\ 
        \bottomrule
    \end{tabular}
\end{table}

As only the training-a dataset contains synchronised PCG and ECG for measuring the OOD performance,
    a PCG-only model will also be trained and used to be evaluated on training-b-f datasets whilst
    the PCG and ECG input model will be evaluated on the SMPECG dataset.

\section{Results}

These results are to demonstrate the performance improvement observed in deep learning models when training is conducted on the augmented dataset. We do not aim to evaluate the performance of the convolutional neural network.

\subsection{In-distribution Performance}

The ID results are for the datasets on which the models were trained.
This shows the increase in performance when training on the augmented dataset compared to
    the original dataset.
As the only dataset being trained on was training-a, these are the only models presented for
    in-distribution performance. \Cref{tab:ID_orig} displays the ID performance when the models are
    trained on the original dataset, with \Cref{tab:ID_aug} displaying the ID performance for models
    trained on the augmented dataset.

\begin{table*}[htbp]
    \caption{\label{tab:ID_orig}Models performance ID trained on the original dataset.}
    \centering
    \scalebox{\resultsscale}
    {\begin{tabular}{lllllllllll}
            \toprule
            \textbf{Dataset} & \textbf{Data} & \textbf{Acc} & \textbf{Acc-mu} & \textbf{TPR} & \textbf{TNR} & \textbf{PPV} & \textbf{NPV} & \textbf{$\text{F1}^{+}$} & \textbf{$\text{F1}^{-}$} & \textbf{MCC}\\[0pt]
            \midrule
            training-a & PCG+ECG & 90.10\% & 89.40\%  & 91.20\% & 87.50\% & 94.50\% & 80.80\% & 92.90\% & 84.0\% & 0.770 \\[0pt]
            training-a & PCG & 70.40\% & 56.00\% & 91.20\% & 20.80\% & 73.20\% & 50.00\% & 81.20\% & 29.40\% & 0.167 \\[0pt]
            \bottomrule
    \end{tabular}}
\end{table*}

\begin{table*}[hbt!]
    \caption{\label{tab:ID_aug}Models performance ID trained on the augmented dataset.}
    \centering
    \scalebox{\resultsscale}
    {\begin{tabular}{lllllllllll}
            \toprule
            \textbf{Dataset} & \textbf{Data} & \textbf{Acc} & \textbf{Acc-mu} & \textbf{TPR} & \textbf{TNR} & \textbf{PPV} & \textbf{NPV} & \textbf{$\text{F1}^{+}$} & \textbf{$\text{F1}^{-}$} & \textbf{MCC}\\[0pt]
            \midrule
            training-a & PCG+ECG & 92.60\% & 93.50\% & 91.20\% & 95.80\% & 98.10\% & 82.10\% & 94.50\% & 88.50\% & 0.836 \\[0pt]
            training-a & PCG  & 84.00\% & 80.20\% & 89.50\% & 70.80\% & 87.90\% & 73.90\% & 88.70\% & 72.30\% & 0.611 \\[0pt]
            \bottomrule
    \end{tabular}}
\end{table*}

\subsection{Out-of-distribution Performance}

The out-of-distribution results are for the datasets the models were not trained on.
Hence, this shows an increase in the generalisation of the models to other datasets that were not
    trained on.
As the dataset being trained on was training-a, all other datasets are presented for the
    out-of-distribution performance. \Cref{tab:OOD_orig} shows the OOD performance on the original
    dataset, with \Cref{tab:OOD_aug} showing the OOD performance when trained on the augmented
    dataset.

\begin{table*}[htbp]
    \caption{\label{tab:OOD_orig}Models performance in OOD trained on the original dataset.}
    \centering
    \scalebox{\resultsscale}
    {\begin{tabular}{lllllllllll}
            \toprule
            \textbf{Dataset} & \textbf{Data} & \textbf{Acc} & \textbf{Acc-mu} & \textbf{TPR} & \textbf{TNR} & \textbf{PPV} & \textbf{NPV} & \textbf{$\text{F1}^{+}$} & \textbf{$\text{F1}^{-}$} & \textbf{MCC}\\[0pt]
            \midrule
            training-b & PCG & 22.90\% & 50.7\% & 99.00\% & 2.30\%  & 21.5\% & 90.00\% & 35.30\% & 4.5\% & 0.040\\[0pt]
            training-c & PCG & 74.20\% & 47.90\% & 95.80\% & 0.00\% & 76.70\% & 0.00\% & 85.20\% & NaN & -0.099\\[0pt]
            training-d & PCG& 49.10\% & 48.50\% & 82.10\% & 14.80\% & 50.0\% & 44.40\% & 62.20\% & 22.20\% & -0.041 \\[0pt]
            training-e & PCG& 40.90\% & 65.80\% & 96.20\% & 35.50\%  & 12.70\% & 99.00\% & 22.50\% & 52.20\% & 0.192\\[0pt]
            training-f & PCG& 52.60\% & 58.60\% & 73.50\% & 43.80\% & 35.70\% & 79.50\% & 48.10\% & 56.50\% & 0.162\\[0pt]
            SMPECG & PCG+ECG  & 56.20\%  & 50.20\% & 98.30\% & 2.20\% & 56.30\% & 50.00\% & 71.60\% & 4.20\% & 0.017 \\[0pt]
            SMPECG & PCG  & 56.20\% & 50.20\% & 98.30\% & 2.20\% & 56.30\% & 50.00 & 71.60\% & 4.20\% & 0.017 \\[0pt]
            \bottomrule
    \end{tabular}}
\end{table*}

\begin{table*}[hbt!]
    \caption{\label{tab:OOD_aug}Models performance in OOD trained on the augmented dataset.}
    \centering
    \scalebox{\resultsscale}
    {\begin{tabular}{lllllllllll}
            \toprule
            \textbf{Dataset} & \textbf{Data} & \textbf{Acc} & \textbf{Acc-mu} & \textbf{TPR} & \textbf{TNR} & \textbf{PPV} & \textbf{NPV} & \textbf{$\text{F1}^{+}$} & \textbf{$\text{F1}^{-}$} & \textbf{MCC}\\[0pt]
            \midrule
            training-b & PCG & 33.30\% & 53.10\% & 87.50\% & 18.70\% & 22.50\% & 84.70\% & 35.80\% & 30.60\% & 0.066 \\[0pt]
            training-c & PCG & 83.90\% & 74.70\% & 91.7\% & 57.10\% & 88.00\% & 66.70\% & 89.80\% & 61.50\% & 0.517\\[0pt]
            training-d & PCG& 52.70\% & 52.00\% & 92.90\% & 11.10\% & 52.00\% & 60.00\% & 66.70\% & 18.80\% & 0.069 \\[0pt]
            training-e & PCG& 84.00\% & 86.00\% & 88.50\% & 83.50\% & 34.50\% & 98.70\% & 49.60\% & 90.50\% & 0.489\\[0pt]
            training-f & PCG& 73.70\% & 60.10\% & 26.50\% & 93.80\% & 64.30\% & 75.00\% & 37.50\% & 83.30\% &0.282\\[0pt]
            SMPECG & PCG+ECG  & 61.90\%  & 57.00\% & 96.60\% & 17.40\% & 60.00\% & 80.00\% & 71.40\% & 28.60\% & 0.237 \\[0pt]
            SMPECG & PCG  & 57.10\% & 51.60\% & 96.60\% & 6.50\% & 57.00\% & 60.00\% & 71.70\% & 11.80\% & 0.073 \\[0pt]
            \bottomrule
    \end{tabular}}
\end{table*}

\newpage
\subsection{{Synthetic Generation Output}}
\begin{figure}[H]
    \centering
    \includegraphics[width=0.8\linewidth]{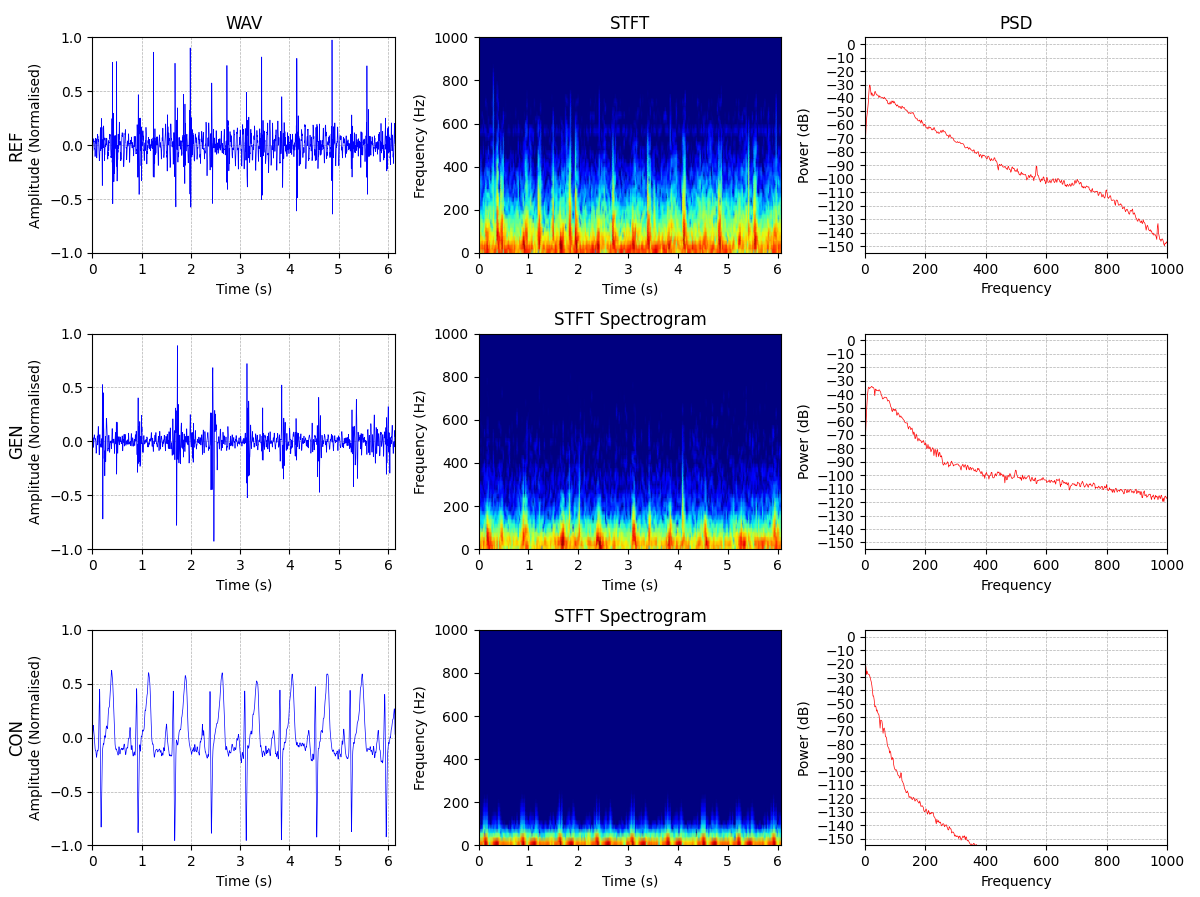}
    \caption{{Generated DiffWave signal.}}
    \label{fig:diff-output}
\end{figure}

\begin{figure}[H]
    \centering
    \includegraphics[width=0.8\linewidth]{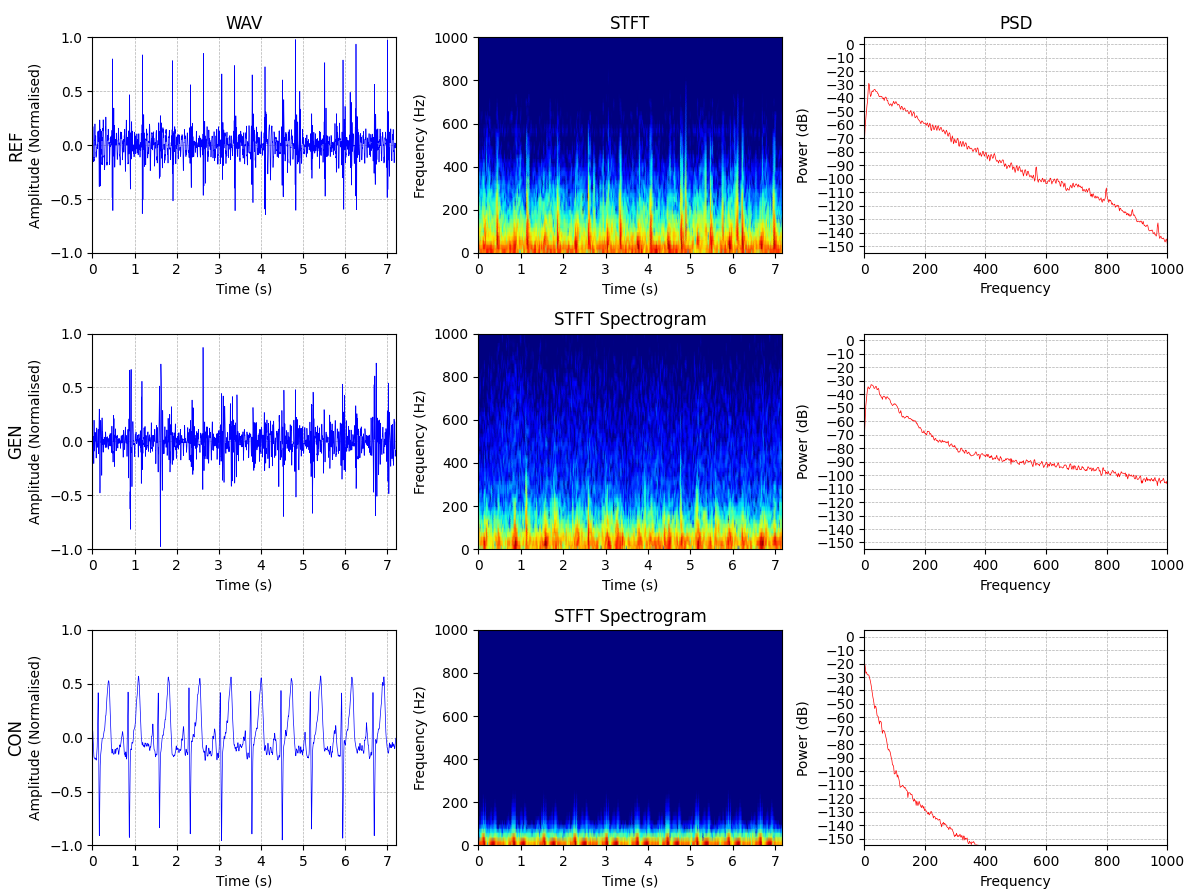}
    \caption{{Generated WaveGrad signal.}}
    \label{fig:wave-output}
\end{figure}

\begin{figure}[H]
    \centering
    \includegraphics[width=0.8\linewidth]{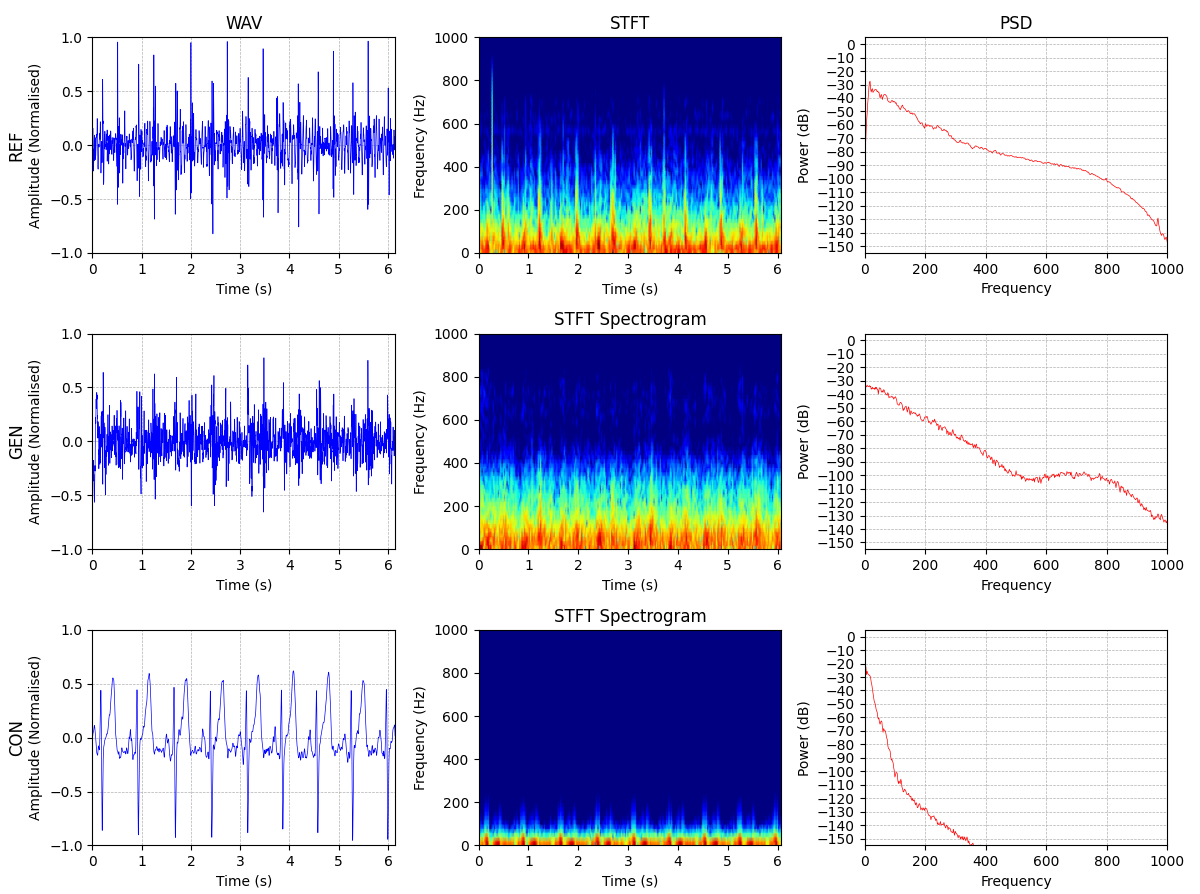}
    \caption{{Generated c$\beta$-VAE signal.}}
    \label{fig:vae-output}
\end{figure}

\begin{figure}[H]
    \centering
    \includegraphics[width=0.8\linewidth]{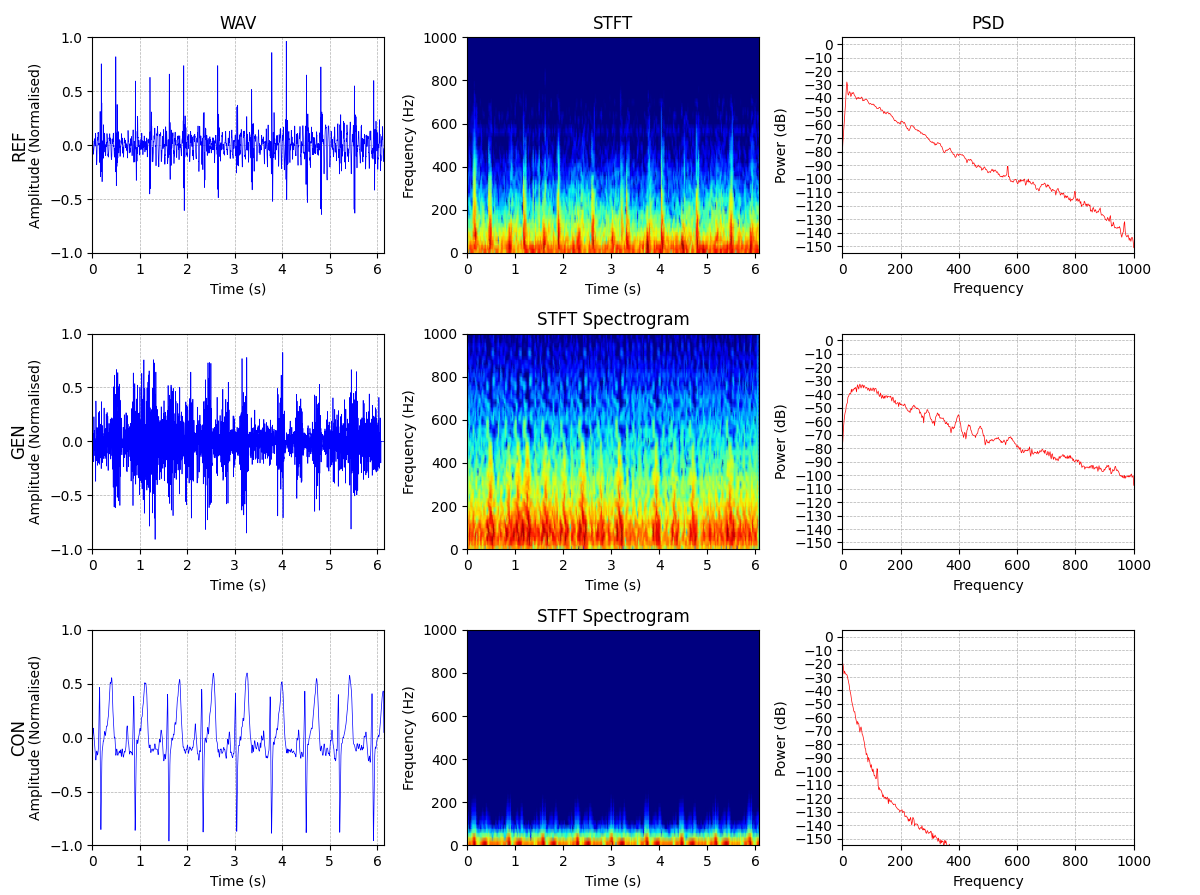}
    \caption{{Generated cDCGAN signal.}}
    \label{fig:dcgan-output}
\end{figure}

{Figure \ref{fig:diff-output} shows the generation output of the DiffWave diffusion model, with the other diffusion models, WaveGrad, output in Figure \ref{fig:wave-output}. The reference signal is from patient `a0040` from the training-a dataset. It is noted that this generation is done from random noise and the conditioning signal, so it will not be reconstructed to look identical to the reference signal. However, it should be time-aligned with the conditioning signal. To better demonstrate the improved generation of the proposed diffusion model, Figure \ref{fig:vae-output} and Figure \ref{fig:dcgan-output} show the outputs from a conditional $\beta$ variational autoencoder (c$\beta$-VAE) and a conditional deep convolutional generative adversarial network (cDCGAN).
Appendices B and C contain the architectures and training used for the c$\beta$-VAE and cDCGAN models, respectively.
It was found that the cDCGAN could not be trained to generate realistic PCG signals as it suffered from mode collapse and struggled with generation. The c$\beta$-VAE resulted in a noisier generated PCG signal than the diffusion model. For this reason, only diffusion models were utilised for the synthetic data generation. These results are inline with the generative trilemma as mentioned in Section \ref{sec:tri}. Further synthetic signals of the diffusion models can be found in \ref{ap:examples}.}

\section{Discussion}

It was found that the ID performance was improved for all models tested, with a 2.5\% improvement in
    accuracy for the PECG model and a 13.6\% improvement in subject-level accuracy for the PCG
    model. 
    The augmented dataset is also shown to improve the balanced accuracy and hence help to balance
    between sensitivity and specificity, with all these being improved from the original dataset to
    the augmented dataset.
This was observed through a balanced accuracy improvement of 4.1\% and 24.2\% for the PECG and PCG
    models, respectively.
    This is further shown by an increase in the MCC value from 0.77 to 0.836 and 0.167 to 0.611 for
    the EPCG and PCG models, respectively.
This shows that by augmenting the original data as well as adding synthetic data, and ensuring a
    balanced dataset, the ID performance can be improved.

The OOD performance was also seen to improve with the augmented dataset.
Although the models were not trained on these datasets, the introduction of augmented data improved
    all model's accuracy and overall robustness, as seen by the increase in MCC values
    across all datasets.
In particular, in the CinC datasets, there was an improvement in accuracy of at most 43.1\% in
    training-e and of at least 3.6\% in training-d, with the improvement in accuracy in all other
    CinC datasets are between these values.
    Further, the balanced accuracy in all of these datasets was improved.
With the greatest increase in balanced accuracy of 26.8\% from training-c and the smallest being 1.5\% from training-f.
    The MCC was also seen to increase in all cases, with the greatest increase of 0.616 occurring in
    training-c and the smallest increase of 0.026 in training-b.
With all performance metrics increasing, the OOD performance was improved by the use of this
    augmented dataset, which shows that these augmentations help to improve the robustness of models
    when used on unseen OOD data.

In the SMPECG dataset, there was a much smaller improvement in accuracy, with an increase of 5.7\%
    with the EPCG model and an increase of 0.9\% with the PCG model.
    Also, balanced accuracy for both models increases, with 6.8\% and 1.4\% for the EPCG
    and PCG models, respectively.
    However, there was a much greater improvement in MCC and overall balancing the performance with
    an increase to the MCC value of 0.22 for the EPCG model and 0.056 for the PCG model.
This shows that although a small improvement, this augmentation helps not only improve
    classification accuracy but also helps to balance the classifier, improving its balanced
    accuracy and MCC values.

As shown, both the ID and OOD performance have been increased by utilising the augmented data,
    achieving the objective of improving the robustness of the classifier.
Better results are found for PCG-only models.
This, however, is due to more data to test with
    than synchronised PCG and ECG data.
However, the OOD for some datasets is still low, showing that there is still room for
    improvement in making a truly robust and general abnormal heart sound classifier.
Utilising a larger dataset and applying these methods, the classifier is expected to become much
    more general, as seen with classifiers trained on this smaller dataset.
{The training datasets and the CinC testing datasets are also not as representative of real-world datasets as compared to the SMPECG dataset; however, with the improvement seen on this dataset, it suggests that when used entirely on real-world datasets, this method will still result in improvement.}

\section{Conclusion and Further Work}

Increasing training data through augmentation has improved ID and OOD performance in classifying abnormal heart sounds.
The use of diffusion models to generate synthetic heart sounds conditioned on ECG signals has successfully enabled the generation of synchronised PCG from ECG data, expanding the data distribution and enhancing classifier robustness.
This is not limited to classifiers that utilise multimodal PCG and ECG data but also for single-mode classifiers that utilise only PCG, as found from the increase in performance and robustness of PCG-only models.
Future work should scale this approach to multichannel PCG signals for use with classifiers that utilise such data.

This study provides evidence that data augmentation, specifically through DDPMs, can significantly enhance the robustness and generalisation of classifiers for abnormal heart sound detection.
By conditioning synthetic PCG signals on ECG data, we generated augmented datasets that improved performance in both ID and OOD scenarios, consistently observed across key metrics such as accuracy, balanced accuracy, and MCC.

Our approach increases the size of training datasets and enriches data diversity, which is crucial for developing models resilient to variations in real-world clinical settings.
The augmentation process effectively addresses data imbalance and noise, providing a stronger foundation for training machine learning models.

However, while the introduced augmentation techniques have shown promise, certain limitations remain, particularly in generalising models to new datasets.
The OOD performance, though improved, suggests that further refinement of these methods is necessary.
This could involve optimising diffusion model parameters or exploring alternative generative approaches that better capture the complex patterns in biomedical signals.

Future work should focus on scaling these methods to accommodate multichannel PCG data, enabling more comprehensive heart sound analysis and potentially improving classification accuracy.
{This will allow the training and test datasets to use data from the SMPECG dataset, demonstrating the effectiveness of this methodology on a real-world dataset.}
This study demonstrates a viable strategy for enhancing classifier performance through synthetic data generation, contributing to more reliable cardiovascular disease diagnosis.

\section*{Authors' contribution}
Leigh Abbott: conceptualisation, methodology, software, formal analysis, validation investigation-data collection, writing-original draft, writing-review \& editing, visualisation.
Milan Marocchi: software, validation, writing-original draft, writing-review \& editing, visualisation.
Matthew Fynn:  writing-review \& editing.
Yue Rong: resources, project administration, writing-review \&
editing, supervision.
Sven Nordholm:
resources, project administration, writing-review \& editing, supervision. 

\section*{Ethics approval and consent}
The study received approval from the ethics committee of Fortis Hospital, Kolkata, India, where the multichannel data collection occurred. Informed consent was obtained from all participating subjects.
All other datasets are open-access, so no approval is required.

\section*{Acknowledgement}
We thank \textit{Ticking Heart Pty Ltd} for allowing the use of data collected from their vest design. We also thank Harry Walters for his valuable remarks and feedback on this work.

\section*{Conflict of Interest}
We declare that we have no conflicts of interest.

\appendix
\section{{Conditional Denoising Diffusion Probabilistic Model equations}}
\label{ap:diff}

{DDPMs considers the conditional distribution $p_{\theta}(\bold{y}_0 | \bold{x)}$, with $\bold{y}_0$ being the
    original waveform and $\bold{x}$ the conditioning features that correspond with $\bold{y}_0$,}

{
\begin{equation}
    \label{eqn:likelihood}
    p_{\theta}\left(\bold{y}_0 | \bold{x}\right) = \int p_{\theta}\left(\bold{y}_{0:N} | \bold{x}\right) d\bold{y}_{1:T}
\end{equation}}

\noindent {where $\bold{y}_1,\ldots{},\bold{y}_T$ is a series of latent variables.
The posterior $q\left(\bold{y}_{1:T} | \bold{y}_0\right)$ is the forward diffusion process, which is
    defined through the Markov chain:}

{
\begin{equation}
    q\left(\bold{y}_{1:T}|\bold{y}_0\right) = \prod_{t=1}^T q\left(\bold{y}_t | \bold{y}_{t-1}\right)
\end{equation}}

{Gaussian noise being added in each iteration is defined as,}

{
\begin{equation}
    q\left(\bold{y}_{t}|\bold{y}_{t-1}\right) = \mathcal{N}\left(\bold{y}_t; \sqrt{1 - \beta_t}\bold{y}_{t-1},\beta_t I\right)
\end{equation}}

\noindent {with the noise being defined with a fixed noise schedule for $\beta_1,\ldots{},\beta_T$.
Hence, the diffusion process can be computed for any $t$ as}

{
\begin{equation}
    \bold{y}_t = \sqrt{\overline{\alpha}_t}\bold{y}_0 + \sqrt{1- \overline{\alpha}_t}\epsilon_t
\end{equation}}

\noindent {where $\alpha_t = 1 - \beta_t$ and $\overline{\alpha_t} = \prod_{i=1}^t \alpha_i$.
As the likelihood in \Cref{eqn:likelihood} is intractable, training these models is done by
    maximising its variational lower bound (ELBO).
\citeauthor{ho2020denoising} found that using a loss as defined in \Cref{eqn:gen_liss} leads to
    higher quality generation.}

{
\begin{equation}
    \label{eqn:gen_liss}
    \E_{t, \epsilon}\left[ \norm{\epsilon_{\theta}\left(\bold{y}_t , \bold{x}, t\right) - \epsilon_t}_2^2 \right]
\end{equation}}

{The model estimates the noise added in the forward process, which is written as $\epsilon_{\theta}$
    and the actual noise added is written as $\epsilon_t$, where $\epsilon_t \sim \mathcal{N}\left(0,
    I\right)$.}

{Generation is then done by first sampling $\bold{y}_{T} \sim \mathcal{N}\left(0,I\right)$ and $\bold{z} \sim
    \mathcal{N}\left(0, I\right)$, before following the below equation until for $t=T,\ldots{},1,0$,}

{\begin{equation}
    \bold{y}_{t-1} = \frac{1}{\sqrt{\alpha_t}}\left( \bold{y}_t - \frac{1 - \alpha_t}{\sqrt{1 - \overline{\alpha}_t}}
        \epsilon_{\theta}\left(\bold{y}_t,\bold{x},t\right)\right) + \sigma_t \bold{z}
\end{equation}}

{\noindent where $\sigma_t = \tilde{\beta_t}$ and $\tilde{\beta_t} = \frac{1 - \overline{\alpha}_{t-1}}{1 -
    \overline{\alpha}_t}\beta_t$ is the variance at step $t$ for $t > 1$ and $\tilde{\beta_1} =
    \beta_1$.}

\section{{c$\beta$-VAE Architecture and Training}}

\begin{figure}[H]
\centering
\includegraphics[width=1.0\linewidth]{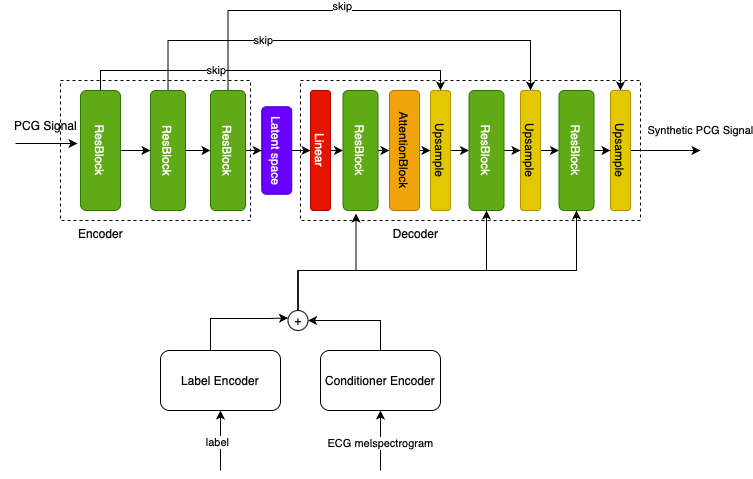}
\caption{{c$\beta$-VAE architecture diagram.}}
\label{fig:vae-arch}
\end{figure}

{The c$\beta$-VAE architecture is shown in Figure~\ref{fig:vae-arch}.
The label and conditioner encoders are single 1D convolutional layers designed to align the label embedding and conditioning features with the feature dimensions of the layers to which they are added.}

{Each residual block (ResBlock) consists of two 1D convolutional layers, each followed by batch normalisation. In the encoder, the first convolution in a ResBlock downsamples the signal via striding, while the second maintains the same number of input and output channels. Both layers use a kernel size of 3; the first layer has a variable stride for downsampling, while the second has a fixed stride of 1.}

{The ResBlocks in the decoder do not perform any upsampling or downsampling. They use the same number of input and output channels for both layers, with all other parameters mirroring those in the encoder. 
The dimensions and sizes of layers are in Table \ref{tab:vae_training}.}

{The model was trained for 24 hours on an RTX 3090 GPU using the same preprocessing steps and collator as the diffusion models. Training used the AdamW optimiser with a learning rate scheduler that reduces the learning rate on loss plateau.}

{The total loss function, shown in Equation~\ref{eq:total_loss}, combines signal reconstruction loss, KL divergence, and multiscale STFT loss. Here, $Z_{\mu}$ and $Z_{\sigma}$ are the encoder’s outputs representing the latent mean and standard deviation, respectively:}

{
\begin{equation}
\mathcal{L} = \beta D_{\text{KL}} \left( \mathcal{N}(Z_{\mu}, Z_{\sigma}) | \mathcal{N}(0, 1) \right) + \alpha \mathcal{L}_{\text{recon}} + \theta \mathcal{L}_{\text{mSTFT}}
\label{eq:total_loss}
\end{equation}}

{The reconstruction loss, $\mathcal{L}_{\text{recon}}$, is defined as the mean squared error (MSE) between the reference signal $\mathbf{s}_{\text{ref}}$ and the generated signal $\mathbf{s}_{\text{gen}}$:}

{
\begin{equation}
\mathcal{L}_{\text{recon}} = \norm{ \mathbf{s}_{\text{ref}} - \mathbf{s}_{\text{gen}} }_2^2
\label{eq:recon_loss}
\end{equation}}

{The multiscale STFT loss, $\mathcal{L}_{\text{mSTFT}}$, is computed by comparing the STFTs of the reference and generated signals using four different window sizes: $w \in\{256, 512, 1024, 2048\}$. Each comparison uses L1 loss, and the results are averaged:}

{
\begin{equation}
\mathcal{L}_{\text{mSTFT}} = \frac{1}{|w|} \sum_{w} \norm{ \mathbf{S}_{\text{ref}, w} - \mathbf{S}_{\text{gen}, w} }_1
\label{eq:stft_loss}
\end{equation}}

{The KL divergence weight $\beta$ was scheduled to increase during training, starting at 0.1 and linearly ramping up to 1.0 over 20,000 steps.}

{The training hyperparameters are summarised in Table~\ref{tab:vae_training}.}

\begin{table}[H]
\centering
\begin{tabular}{>{}l>{}c}
\toprule
\textbf{Hyperparameter} & \textbf{Value} \\
\hline
Initial learning rate & $2 \times 10^{-4}$ \\
Batch size & 8 \\
Latent dimension size & 32 \\
Label embedding dimension size & 32 \\
ResBlock feature sizes (blocks 1–3) & 32, 64, 128 \\
ResBlock strides (blocks 1–3) & 2, 2, 2 \\
$\alpha$ (reconstruction loss weight) & 1.0 \\
$\theta$ (STFT loss weight) & 0.5 \\
\bottomrule
\end{tabular}
\caption{{Training hyperparameters for the c$\beta$-VAE model.}}
\label{tab:vae_training}
\end{table}

\section{{cDCGAN Architecture and training}}
\begin{figure}[H]
\centering
\includegraphics[width=1.0\linewidth]{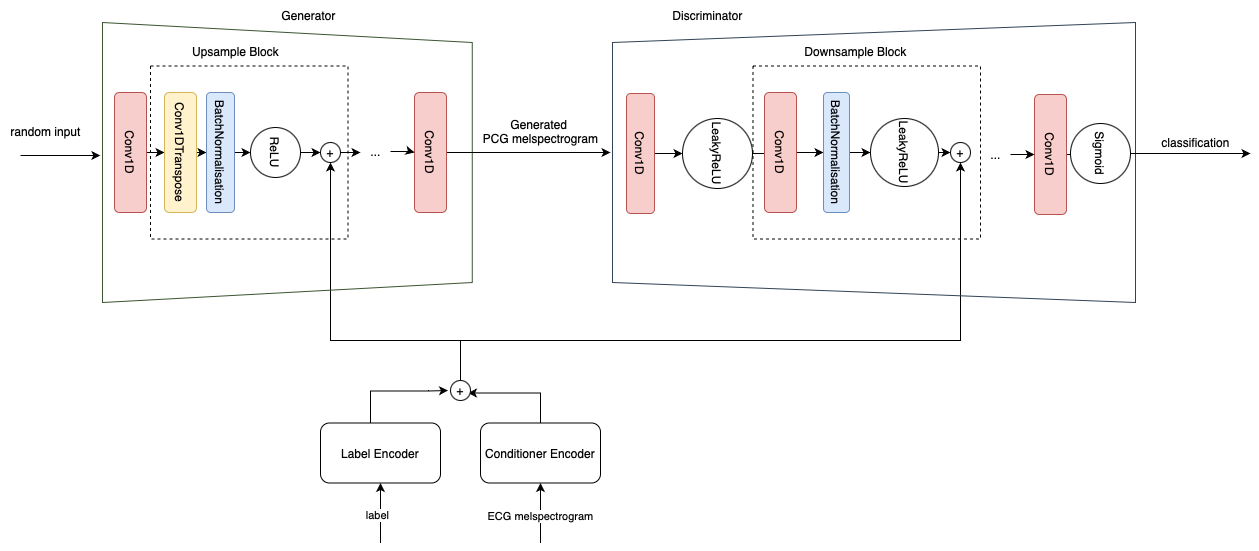}
\caption{{cDCGAN architecture diagram.}}
\label{fig:dcgan-arch}
\end{figure}

{The cDCGAN architecture is shown in Figure \ref{fig:dcgan-arch}. The label and conditioner encoders are single 1D convolutional layers designed to align the label embedding and conditioning features with the feature dimensions of the layers to which they are added. The size of the layers and number of layers are shown in Table \ref{tab:gan-params}.
The model was trained for 24 hours on an RTX 3090 GPU using the same preprocessing steps and collator
as the diffusion models. Training used the AdamW optimiser with a learning rate scheduler that reduces the
learning rate on loss plateau.}

{The total loss function shown in Equation \ref{eqn:gan-loss}, combines the adversarial loss with signal reconstruction loss, multiscale STFT loss and feature matching loss.}

{
\begin{equation}
 \mathcal{L} = \beta\mathcal{L}_{\text{Adv}} + \alpha\mathcal{L}_{\text{recon}} + \theta\mathcal{L}_{\text{mSTFT}} + \lambda\mathcal{L}_{FM}
\label{eqn:gan-loss}
\end{equation}}

{The adversarial loss is defined as the sum of the discriminator and generator losses as in Equation \ref{eqn:adv-loss}.}
{
\begin{equation}
\mathcal{L}_{Adv} = \mathcal{L}_D + \mathcal{L}_G
\label{eqn:adv-loss}
\end{equation}}

{The discriminator loss is described as follows,}
{\begin{equation}
\mathcal{L}_D = -\mathbb{E}_{x \sim p_{\text{data}}(x)} \left[ \log D(x, y) \right] 
               - \mathbb{E}_{z \sim p_z(z)} \left[ \log \left( 1 - D(G(z, y), y) \right) \right]
\end{equation}}

{The generator loss is found in Equation \ref{eqn:gen-loss}.}
{\begin{equation}
\mathcal{L}_G = -\mathbb{E}_{z \sim p_z(z)} \left[ \log D(G(z, y), y) \right]
\label{eqn:gen-loss}
\end{equation}}

{where:
\begin{itemize}
    \item $x \sim p_{\text{data}}(x)$: Real data sampled from the true data distribution.
    \item $z \sim p_z(z)$: Latent vector sampled from Gaussian noise (random input).
    \item $y$: Conditional input (ECG mel spectrogram).
    \item $G(z, y)$: Generator output conditioned on random input $z$ and conditional input $y$.
    \item $D(x, y)$: Discriminator's estimate of the probability that $x$ is real, given condition $y$.
    \item $\mathcal{L}_D$: Discriminator loss—maximized to distinguish real from fake samples.
    \item $\mathcal{L}_G$: Generator loss—minimized to fool the discriminator.
\end{itemize}}

{The reconstruction loss and multiscale STFT loss are the same that were utilised within the c$\beta$-VAE model in Equation \ref{eq:recon_loss} and Equation \ref{eq:stft_loss}.}

{Lastly, the feature matching loss is defined as the L1 loss between the features, embedding taken from the last downsample block, of the discriminator for the reference signal and the generated signal, where $F_{ref}$ are the features of the reference signal $F_{gen}$ are the features of the generated signal.}

{
\begin{equation}
    \mathcal{L}_{FM} = \norm{ F_{ref} - F_{gen} }_1
\end{equation}}

{The training hyperparameters are summarised in Table \ref{tab:gan-params}.}

\begin{table}[H]
\centering
\begin{tabular}{>{}l>{}c}
\toprule
\textbf{Hyperparameter} & \textbf{Value} \\
\hline
Initial learning rate & $2 \times 10^{-4}$ \\
Batch size & 8 \\
Latent dimension (random input) dimension & 32 \\
Generator/Discriminator feature dimension size & 64 \\
Upsample/Downsample scales & 4, 2, 2, 6 \\
$\alpha$ (reconstruction loss weight) & 1.0 \\
$\beta$ (adversarial loss weight) & 1.0 \\
$\theta$ (STFT loss weight) & 5.0 \\
$\lambda$ (feature matching loss weight) & 5.0 \\
\bottomrule
\end{tabular}
\caption{{Training hyperparameters for the cDCGAN model.}}
\label{tab:gan-params}
\end{table}

\section{{Diffusion model generation examples}}
\label{ap:examples}
\begin{figure}[H]
    \centering
    \begin{subfigure}[b]{0.45\linewidth}
        \centering
        \includegraphics[width=\linewidth]{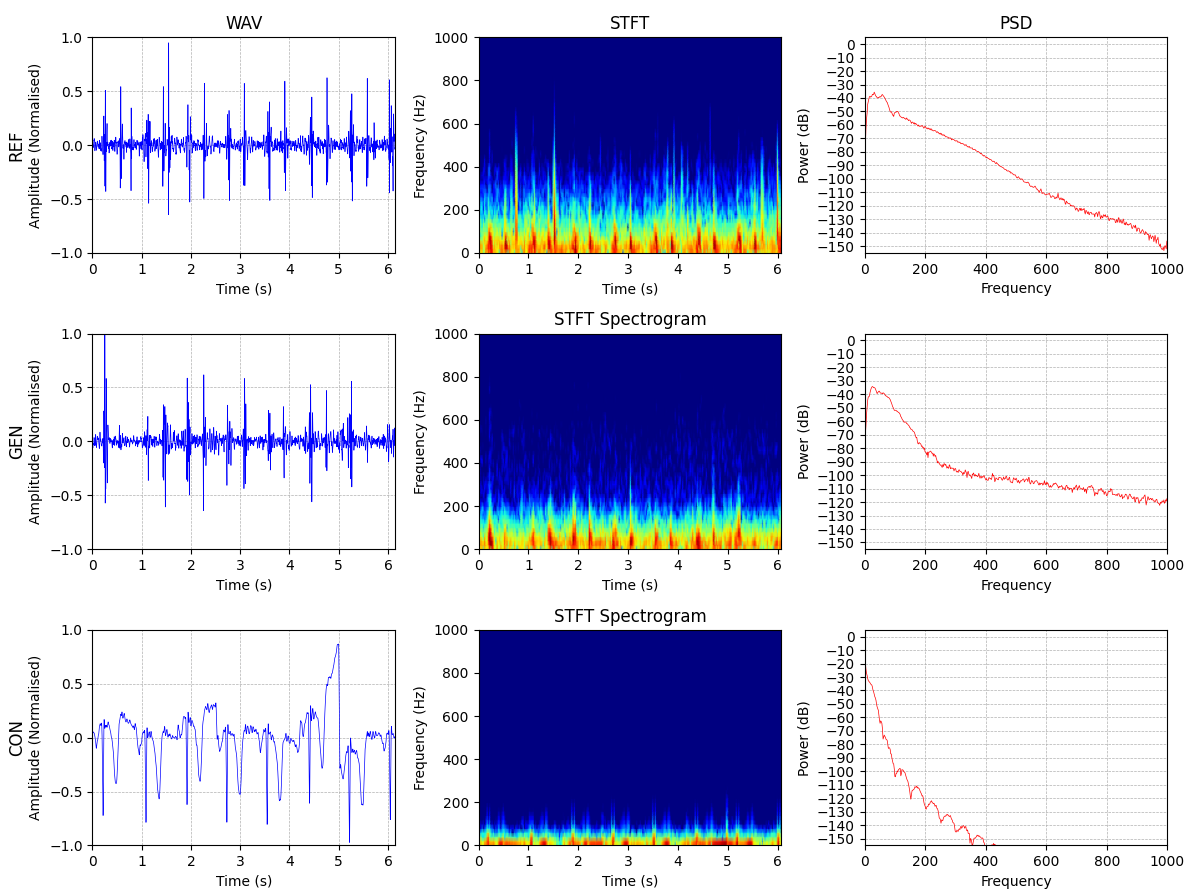}
        \caption{{Patient a0093}}
        \label{fig:diffwave-a0093}
    \end{subfigure}
    \hspace{0.05\linewidth}
    \begin{subfigure}[b]{0.45\linewidth}
        \centering
        \includegraphics[width=\linewidth]{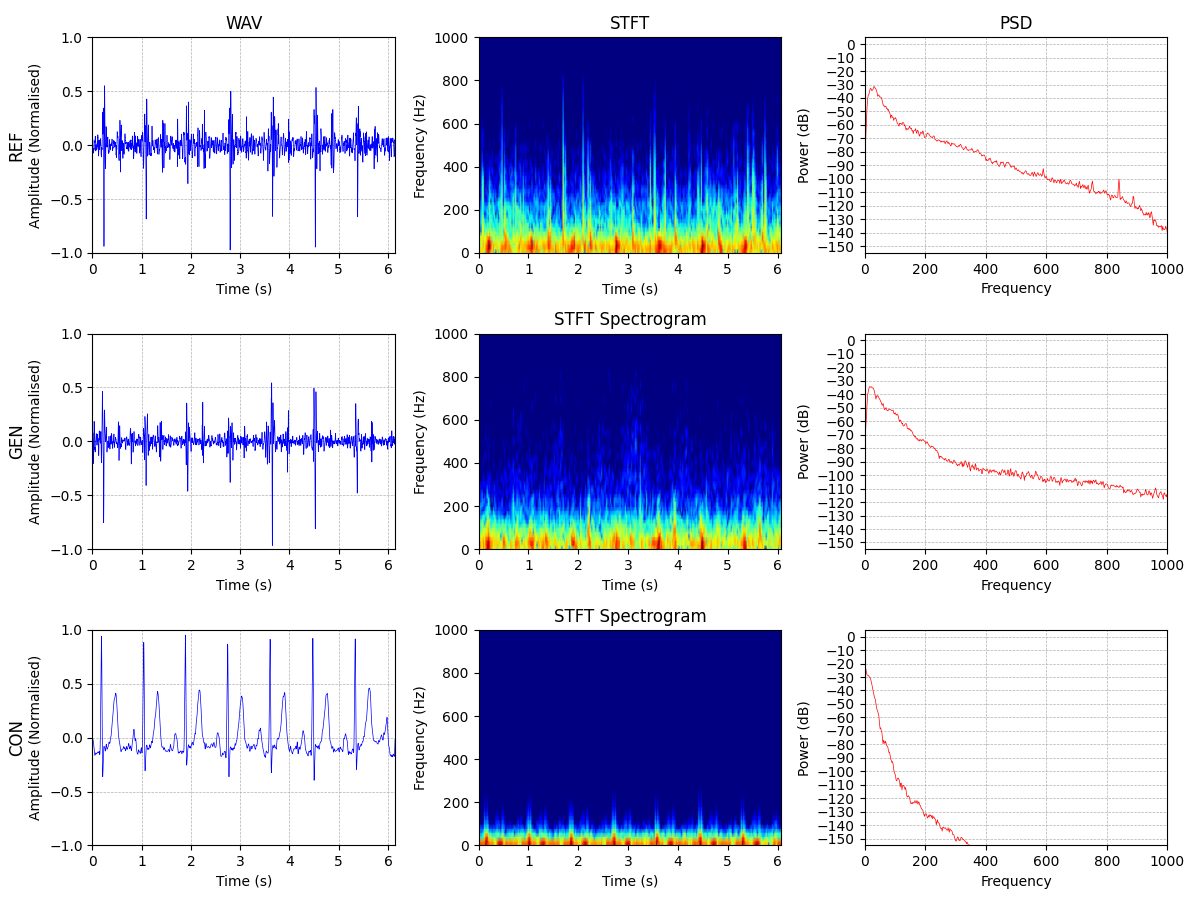}
        \caption{{Patient a0103}}
        \label{fig:diffwave-a0103}
    \end{subfigure}
    \vspace{1em} 
    \begin{subfigure}[b]{0.45\linewidth}
        \centering
        \includegraphics[width=\linewidth]{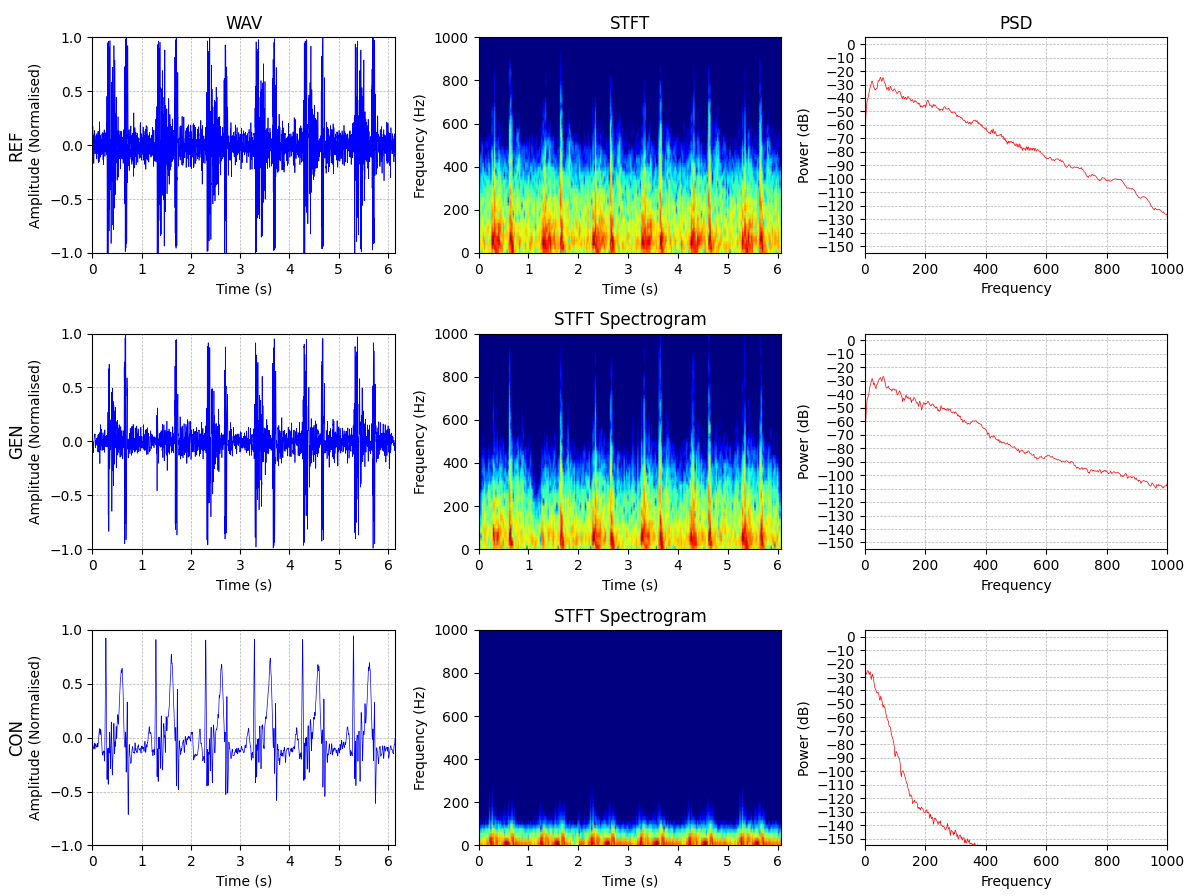}
        \caption{{Patient a0400}}
        \label{fig:diffwave-a0400}
    \end{subfigure}

    \caption{{Generated DiffWave signals for three different patients.}}
    \label{fig:diffwave-subfigures}
\end{figure}

\begin{figure}[H]
    \centering
    \begin{subfigure}[b]{0.45\linewidth}
        \centering
        \includegraphics[width=\linewidth]{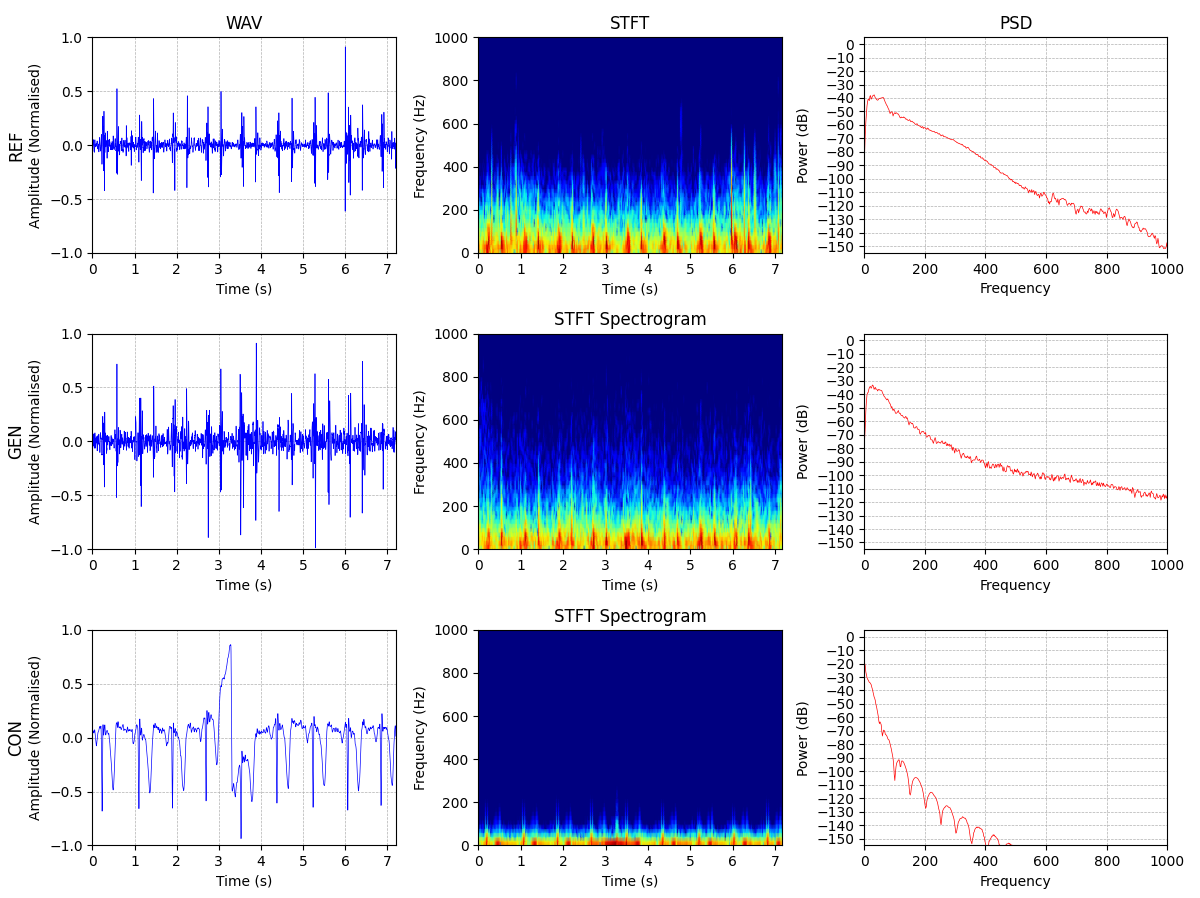}
        \caption{{Patient a0093}}
        \label{fig:wavegrad-a0093}
    \end{subfigure}
    \hspace{0.05\linewidth}
    \begin{subfigure}[b]{0.45\linewidth}
        \centering
        \includegraphics[width=\linewidth]{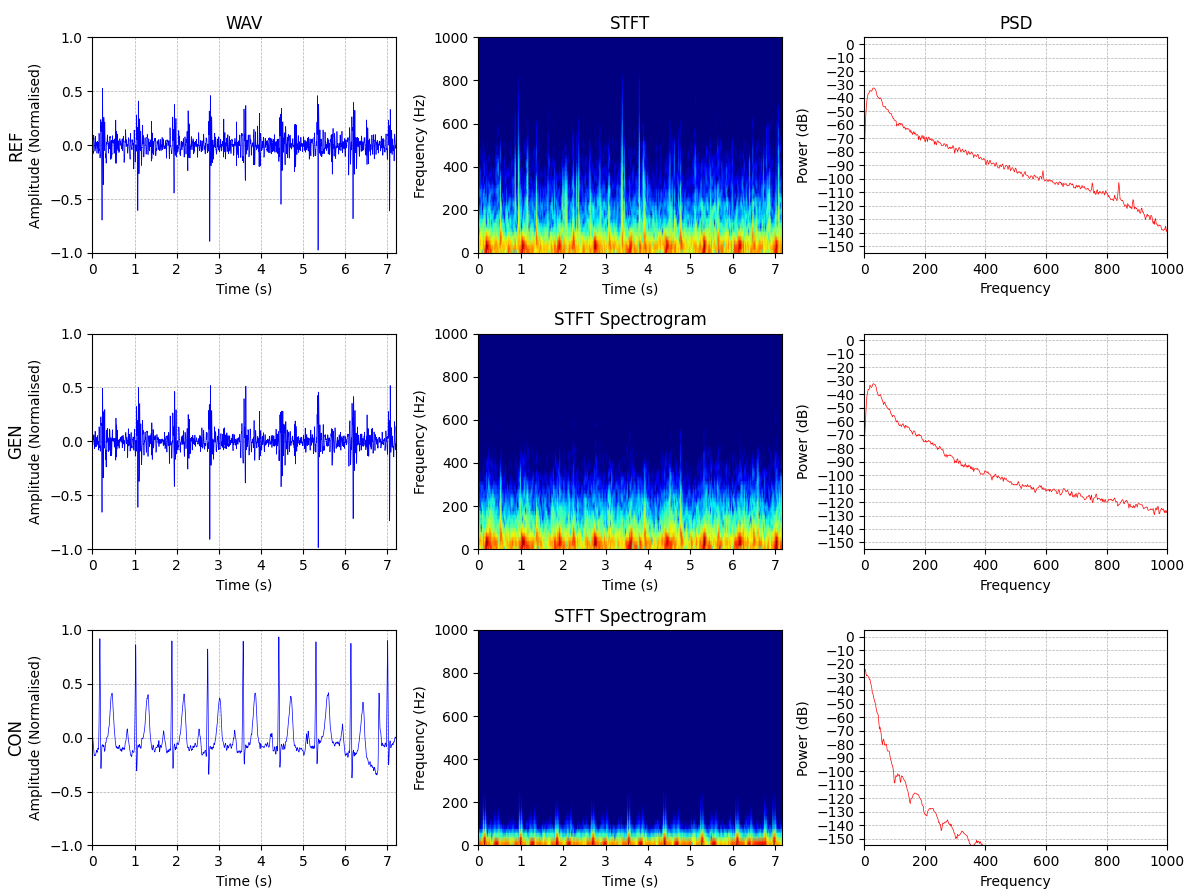}
        \caption{{Patient a0103}}
        \label{fig:wavegrad-a0103}
    \end{subfigure}
    \vspace{1em}
    \begin{subfigure}[b]{0.45\linewidth}
        \centering
        \includegraphics[width=\linewidth]{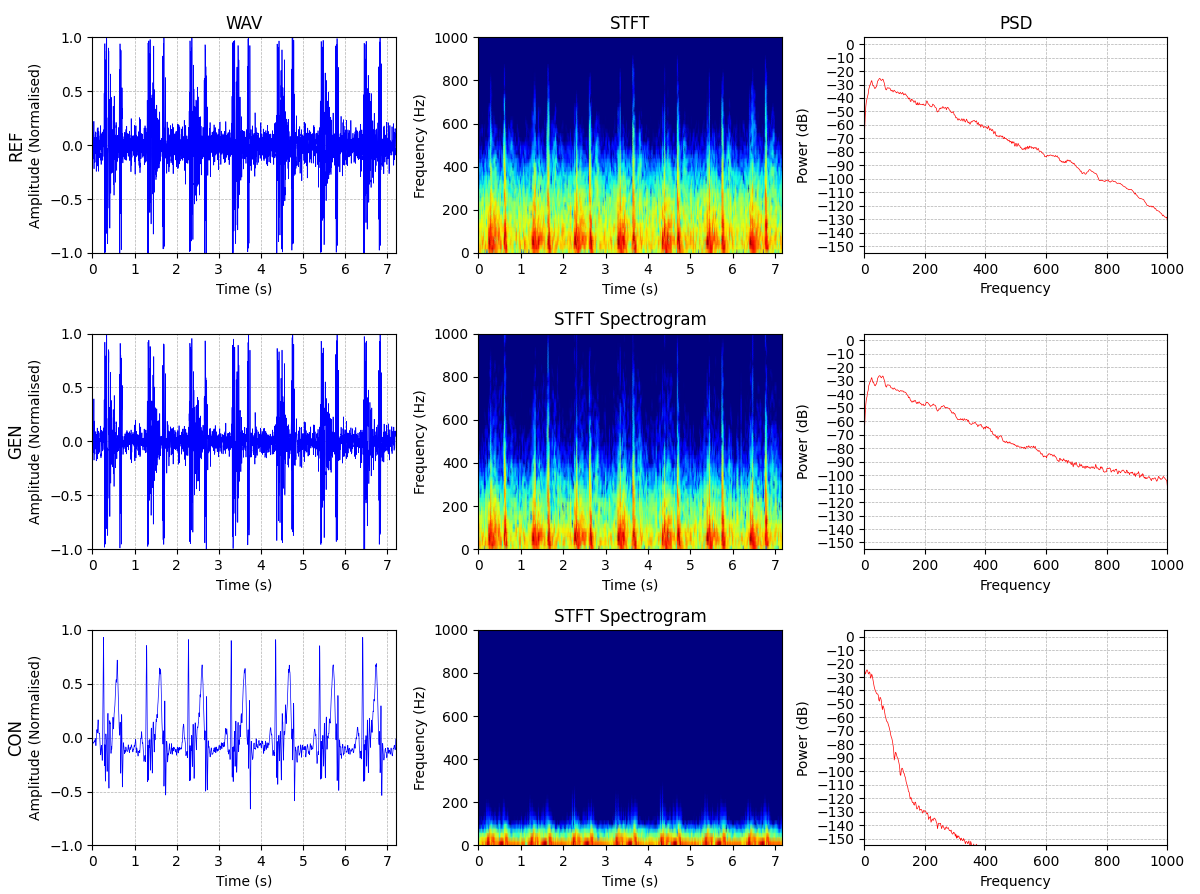}
        \caption{{Patient a0400}}
        \label{fig:wavegrad-a0400}
    \end{subfigure}

    \caption{{Generated WaveGrad signals for three different patients.}}
    \label{fig:wavegrad-subfigures}
\end{figure}

\bibliographystyle{IEEEtranN}
\bibliography{references.bib}

\end{document}